\documentclass[usenatbib]{mn2e}
\usepackage{natbib}
\usepackage{graphicx}
\usepackage{amssymb}
\usepackage{xspace}
\usepackage{epsfig}
\usepackage{times}

\renewcommand\sun{\odot}

\newcommand{\msun}{$M_\sun$\xspace}

\title[Dust formation in clumpy SN shells] {Dust yields in clumpy SN shells: SN~1987A revisited }
\author[Ercolano et al.]{B. Ercolano$^{1,2}$, M. J. Barlow$^1$, B. E. K. Sugerman$^{3,4}$ \\
$^1$Department of Physics and Astronomy, University College London, Gower Street, London WC1E~6BT, UK\\
$^2$Harvard-Smithsonian Centre for Astrophysics, 60 Garden Street, Cambridge, MA 02138, USA\\
$^3${Space Telescope Science Institute, 3700 San Martin Drive, Baltimore,
MD 21218, USA}\\ 
$^4${Goucher College, 1021 Dulaney Valley Rd, Baltimore, MD 21204, USA; ben.sugerman@goucher.edu} 
\\}
\date{Received:}

\begin{document}
\maketitle
 
\begin{abstract}

We present a study of the effects of clumping on the emergent spectral
energy distribution (SED) from dusty supernova (SN) shells illuminated
by a diffuse radiation source distributed throughout the medium.
These models are appropriate for Type~II SNe older than a few
hundred days, when the energy input is dominated by $\gamma$-rays from
$^{56}$Co decay.

The fully 3D radiation transport problem is solved using a Monte
Carlo code, {\sc mocassin}, and we present a set of models aimed
at investigating the sensitivity of the SEDs to various clumping
parameters.  We find that, contrary to the predictions of analytical
prescriptions, the combination of an optical and IR observational data
set is sufficient to constrain dust masses even in the case where
optically thick clumps are present.

Using both smoothly varying and clumped grain density distributions,
we obtain new estimates for the mass of dust condensed by the Type~II
SN~1987A by fitting the optical and infrared spectrophotometric data of
\citet{Woo93} at two epochs (day 615 and day 775). When using
amorphous carbon grains, our best fits to the
observational data imply that about 2.0$\cdot$10$^{-4}$~M$_{\odot}$ of dust had condensed in 
the envelope of SN1987A by day 615 and between 2.0$\cdot$10$^{-4}$ and
4.2$\cdot$10$^{-3}$~M$_{\odot}$ by day 775. We find that the absence of a silicate
emission or absorption feature in the observed mid-IR spectra implies that
no more than 15\% of the dust formed around SN~1987A can have been in the
form of silicate particles. Our models require larger
dust masses for the case of graphite grains, namely between
4.2$\cdot$10$^{-4}$ and 6.6$\cdot$10$^{-4}$~M$_{\odot}$ at day 615 and
between 4.5$\cdot$10$^{-4}$ and 7.5$\cdot$10$^{-4}$~M$_{\odot}$ at day
775.  From our
numerical models we derive dust masses for SN 1987A that are comparable to
previous analytic clumped graphite grain mass estimates, and at least two
orders of magnitude below the 0.1--0.3~M$_{\odot}$ that have been predicted to
condense as dust grains in primordial core collapse supernova ejecta. This low
condensation efficiency for SN~1987A is in contrast to the case of
SN~2003gd, for which a dust condensation efficiency as large as 0.12 has
recently been estimated.

\end{abstract}

\nokeywords


\section{Introduction}

The production of dust in the ejecta of core-collapse supernovae
(CC-SNe) is supported by observations
\citep{Luc89,BD93,Woo93,Elm03,Sug06}, while theoretical studies of
dust condensation in primordial CC-SNe have predicted that 0.1--0.3~\msun of dust
could be produced \citep{Koz91,TF01}. Evidence for dust formation by
(at least) some SNe comes from precise isotopic abundance ratio
studies of grain inclusions found in meteorites, with many (including
graphite and silicon carbide inclusions) exhibiting isotopic
distributions that differ significantly from those found in the Sun
and Earth \citep[e.g.][]{Cla97,Tra99}. However, direct evidence that
SNe play a major role in the dust budget of galaxies is still sparse,
since of the few supernovae that have been probed for dust production,
most appear to have formed much less dust than predicted by
models. For example, analyses of the photometric and spectroscopic
evolution of SNe 1987A and 1999em from 400--800 days after outburst
yielded estimates of a few$\times10^{-4}$ \msun of dust formed per SN
\citep{Luc89,Woo93,Elm03}, which translates into condensation
efficiencies, defined here as (mass of refractory elements condensed
into dust)/(mass of refractory elements in ejecta), of $\sim 10^{-3}$
\citep{Luc89,Woo95}. In contrast, for SNe to have produced the
majority of the $\gtrsim$$10^8$ \msun of dust found in high-redshift
($z>6$) quasars \citep{Ber03,Mai04,Rob04,Hir05}, their condensation
efficiency must have been around $\sim$0.2 \citep{ME03}.

In the cases of both SNe 1987A and 1999em, the above investigators all
separately noted that the dust masses could be significantly larger if
the dust was clumped rather than distributed homogeneously throughout
the ejecta. The presence of clumped dust is supported by the
wavelength-independent circumstellar extinction seen in some SNe
\citep{Luc91,Elm03}, and by the theoretical expectation that
Rayleigh-Taylor instabilities should form in post-shock ejecta
\citep{CK78,HW94}. \citet{Sug06} confirmed this hypothesis for dust
formation within SN~2003gd, by demonstrating with three-dimensional
Monte-Carlo radiative transfer (RT) models that clumpy, inhomogeneous
dust models can require up to an order of magnitude more mass, than
implied by either analytic estimates or by homogeneous dust models, to
produce a given optical extinction and mid-infrared (IR)
excess. Sugerman et al. found that clumped models implied that up to
$2\cdot 10^{-2}$ \msun of dust had formed in the ejecta of SN~2003gd
by day 678, corresponding to a dust condensation efficiency of up
0.12, which is interestingly close to the efficiency of $\sim$0.2
required to match high redshift dusty sources.

Since the previous, rather low, dust mass estimates for SN~1987A were
based on analytical techniques, we were therefore motivated to re-analyse
the available observational data for this object using modelling
techniques similar to those used by Sugerman et al. for SN~2003gd. In
addition to allowing for clumped dust distributions, we also study the
effects that the geometry of the medium and the spatial distribution of
the luminosity source have on the emerging spectral energy distributions
(SED) and, as a consequence, on dust masses determined by fitting RT
models to observations. Radiative transfer in an inhomogeneous medium has
been the subject of a number of studies, including the simplest approach
of a two-phase medium consisting of dense clumps embedded in a less dense
interclump medium (ICM). These studies employed a combination of
analytical techniques and Monte Carlo modelling with dust models of
varying complexity, in order to study dust in different phases of the
interstellar medium (ISM) and in protogalactic and galactic environments
\citep[e.g.][]{Nat84,Boi90,Neu91,HS93,HP93,Wit96,Gor97,Wol98,VD99}.

All the models presented in this work were performed using the 3D Monte
Carlo RT code {\sc mocassin} \citep{Erc03,Erc05}, which can account for
both (primary and secondary) components of the radiation field in a fully
self-consistent manner through absorption, re-emission and scattering of
photons. This is necessary if one wants to deal with the transfer through
inhomogeneous clumpy distributions, both for the case of a diffuse
illuminating source distributed through the medium or for the case of a
central illuminating source. Section~2 describes the basic assumptions
used for the modelling, while the results of our parameter investigation
are presented and discussed in Section~3. In Section~4, we readdress the
issue of the mass of dust formed by SN~1987A, constraining our models
primarily with the observations between $\sim$0.3\,$\mu$m and
30\,$\mu$m published by \citet[][hereafter W93]{Woo93}.

\section{Modelling Strategy}
\label{sec:modstra}

\begin{figure}
\begin{center}
\begin{minipage}[t]{8.0cm}
\includegraphics[width=8.0cm]{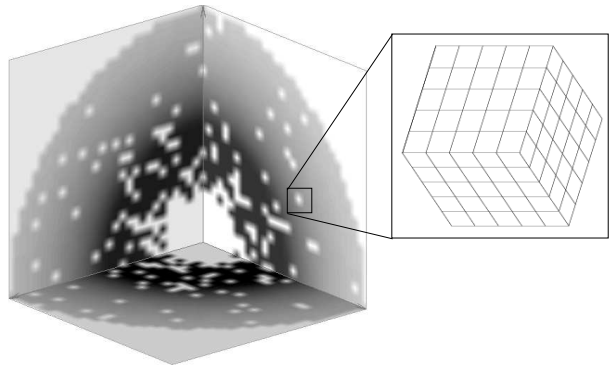}
\end{minipage}
\caption[]{A 3D representation of the dust number density in clumpy 
models.}
\label{fig:clumpy}
\end{center}
\end{figure}

Using both smoothly varying and clumped grain density distributions,
the 3D Monte Carlo radiative transfer (RT) code {\sc mocassin}
\citep{Erc03,Erc05} was used by \citet{Sug06} to derive the mass of
dust condensed by the Type~II SN~2003gd, by fitting optical (B, V, R,
I) and {\it Spitzer Space Telescope} 3.6$\mu$m, 4.5$\mu$m, 5.8$\mu$m,
8.0$\mu$m, and 24$\mu$m fluxes observed at two epochs (days 499 and
678 after outburst).  Their results clearly showed that analytical
analyses of optical and IR data, as well as RT models with smoothly
varying dust density distributions, can underestimate the dust mass by
an order of magnitude or more compared to models that allow for
clumping. \citet{luc89,luc91} provided evidence that optically thick clumps may
exist in the envelope of SN~1987A, implying that the dust mass may be
underestimated by SED-based smooth density distribution models. These
authors argued that amorphous part dust was required by the
wavelength-dependent blueward shift of various emission lines, while
clumps are required to explain the wavelength independent reddening effect
of the dust. Successively, 
W93 summarised the evidence for 
significant clumping in the ejecta of SN~1987A, which was provided by 
observations ranging from $\gamma$-ray and X-ray to infrared
wavelengths.  
We have
therefore revisited the observational data for SN~1987A, employing a
similar modelling strategy to that used for SN~2003gd, i.e.\ we assume
that dust condensed in clumps within the ejecta, with the heating
source distributed within the remnant shell.  Our models are
appropriate for Type~II SNe older than a few hundred days, when
the energy input is dominated by $\gamma$-rays from $^{56}$Co decay
\citep[][W93]{BD93}. We consider two clumping models, designed to
represent the extreme cases of likely clumping behaviour: ({\sc i}) the
diffuse heating source is mixed with the dust grains, which are
present in both the clumps and in the interclump medium (ICM); and
({\sc ii}) the diffuse illumination source is confined to the
virtually dust-free ICM, with all dust grains confined to the
clumps. The latter model appears the most physically plausible, given
that the outer C, O, and Si-rich shells, where carbonaceous and
silicate grains are expected to form, should never completely mix with
the inner $^{56}$Ni/$^{56}$Co zone \citep{Arn89}. Only macroscopic
mixing has been found in the clumpy ejecta of the Type~II supernova
remnant Cas~A \citep{Dou99}, suggesting that the element layers were
not homogeneously mixed.  For completeness, we also consider a
``smooth'', clump-free model.


\citet{Mos89} reported early IR observations (day 265 and 267) and
found no evidence for any emission from dust that might have formed in
the SN ejecta or from pre-existing dust in the surrounding medium.
A variety of observations indicated that the onset of dust formation
in the ejecta of SN~1987A occurred at about day 530 \citep[][
W93]{Luc91}. Given that the featureless 10-$\mu$m spectrum of SN~1987A
after day~530 \citep[W93; ][]{Roc93} precludes a significant silicate
contribution, the dust was assumed to be either amorphous carbon
\citep[optical constants from][]{Han88} or graphite \citep[optical
constants from][]{DL84}. In view of theoretical predictions that
supernova grain size distributions should be relatively lacking in
large particles \citep{TF01}, we adopted a standard \citep[][hereafter
MRN]{MRN77} ISM grain size distribution, N($a$) $\propto a^{-3.5}$,
but with the grain radii $a$ truncated at the upper end, such that
$a_{\rm min}$~=~0.005~$\mu$m and $a_{\rm max}$~=~0.05~$\mu$m.  We
found however that for the models presented in Section~3, the SEDs
predicted by considering a full MRN distribution versus a truncated
size distribution are quite similar and that the choice of either did
not affect the general results of this paper.  Following our parameter
investigation, we determine best fits to the W93 day 615 and day 775
spectra for all three dust configurations (smooth, Clumpy {\sc i} and
Clumpy {\sc ii}) and in Section 4 present revised dust mass estimates
for SN~1987A on days 615 and 775.

\begin{figure}
\begin{center}
\begin{minipage}[t]{8.5cm}
\includegraphics[width=8.5cm]{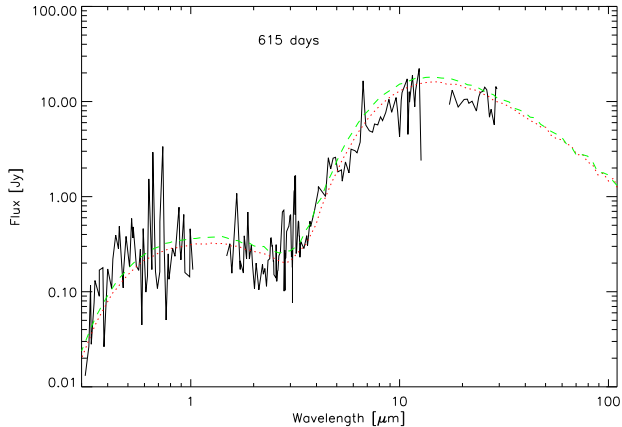}
\end{minipage}
\caption[]{Amorphous carbon model fits to the day 615 spectrophotometry 
published by W93 (black line). Two smooth density models of day
615 with different bolometric luminosity are compared. The red-dotted line is for a
model with $L~=~5.7\times10^5L_{\odot}$ model and the green-dashed line is
for a model with $L~=~6.7\times10^5L_{\odot}$. }
\label{fig:ltest}
\end{center}
\end{figure}

\begin{figure*}
\begin{center}
\begin{minipage}[t]{8.5cm}
\includegraphics[width=8.5cm]{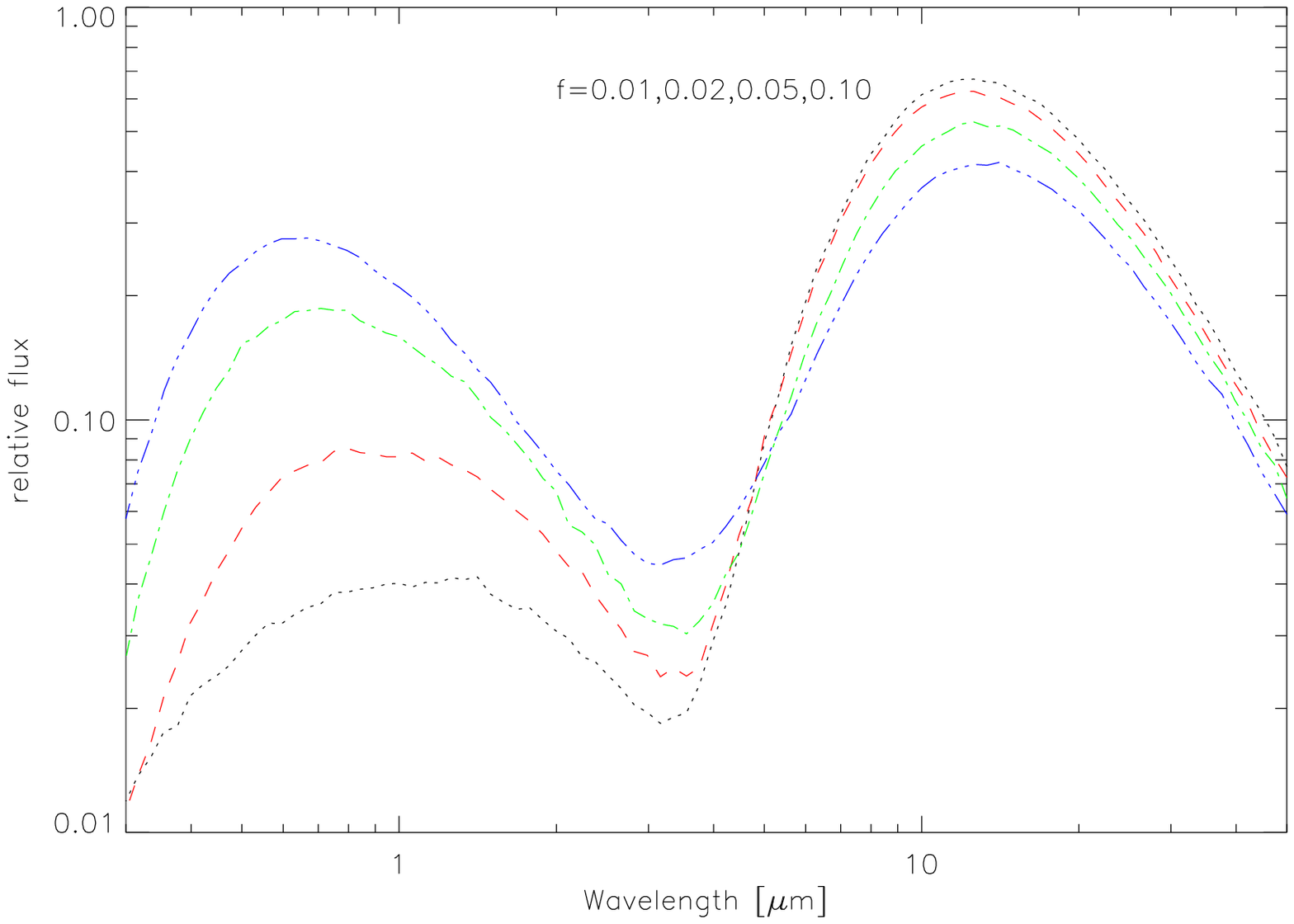}
\end{minipage}
\begin{minipage}[t]{8.5cm}
\includegraphics[width=8.5cm]{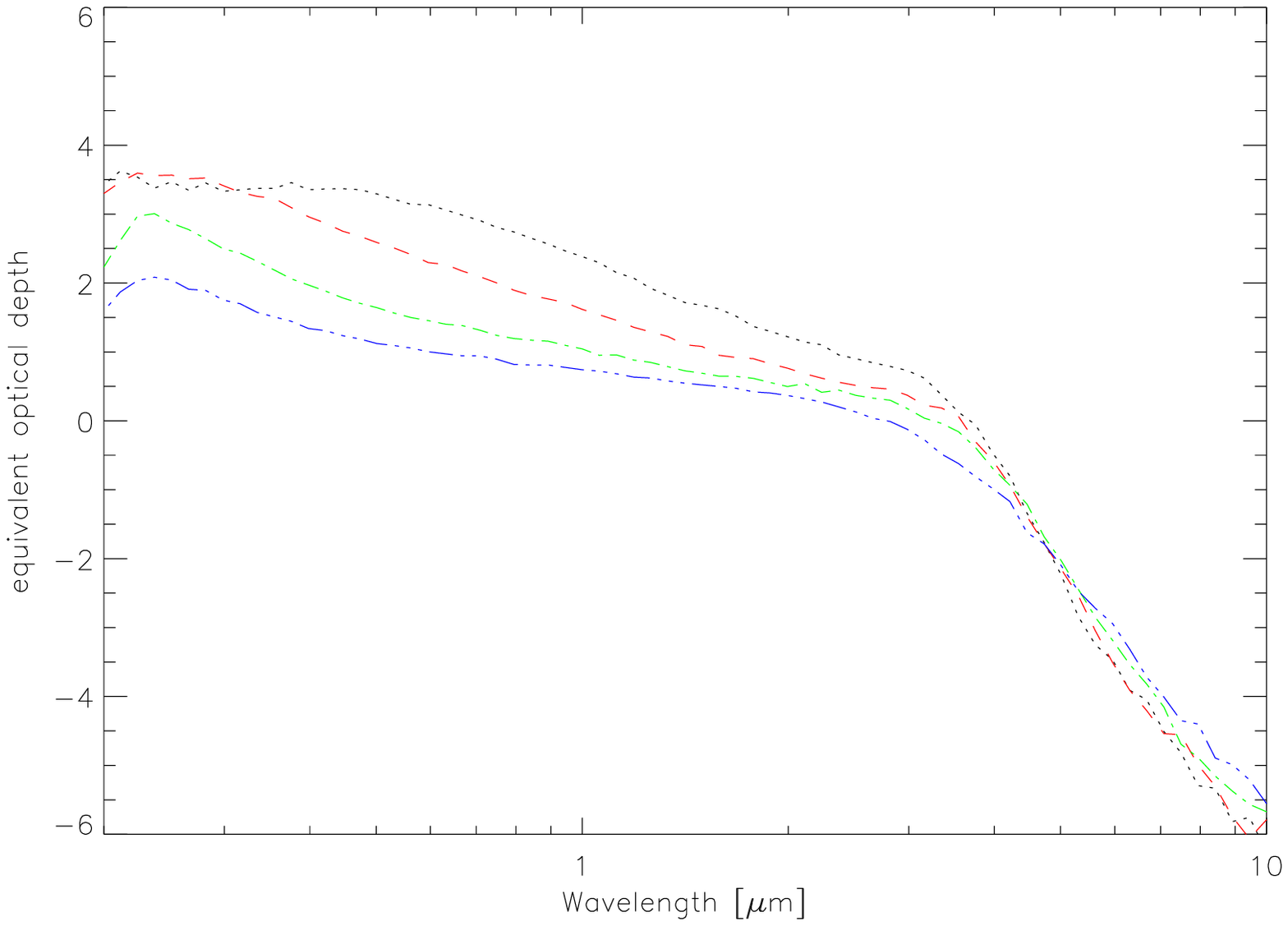}
\end{minipage}
\caption[]{Clumpy~{\sc i} models: dependence of SEDs (left panel) and
equivalent optical depths (right panel) on the volume filling factor,
$f$, of the clumps.  In this and subsequent plots, SEDs (left) are
plotted as dimensionless, distance- and luminosity-independent
spectral shapes, $\lambda\,F_{\lambda}$/$\int{F_{\lambda}d\lambda}$
versus $\lambda$ in microns; equivalent optical depths (right) are
defined by Equation (1), and may thus be negative if the escaping flux
is greater than the incident energy, as is the case in the mid-IR.
All models have the same $M_{\rm d}$~=~5.6$\cdot$10$^{-4}M_{\odot}$
and $\alpha$~=~100.  Colour coding: black (dotted) for $f$~=~0.01, red
(dashed) for $f$~=~0.02, green (dash-dot) for $f$~=~0.05 and blue
(dash-dot-dot) for $f$=0.10. }
\label{fig:sedcef}
\end{center}
\end{figure*}

\begin{figure*}
\begin{center}
\begin{minipage}[t]{8.5cm}
\includegraphics[width=8.5cm]{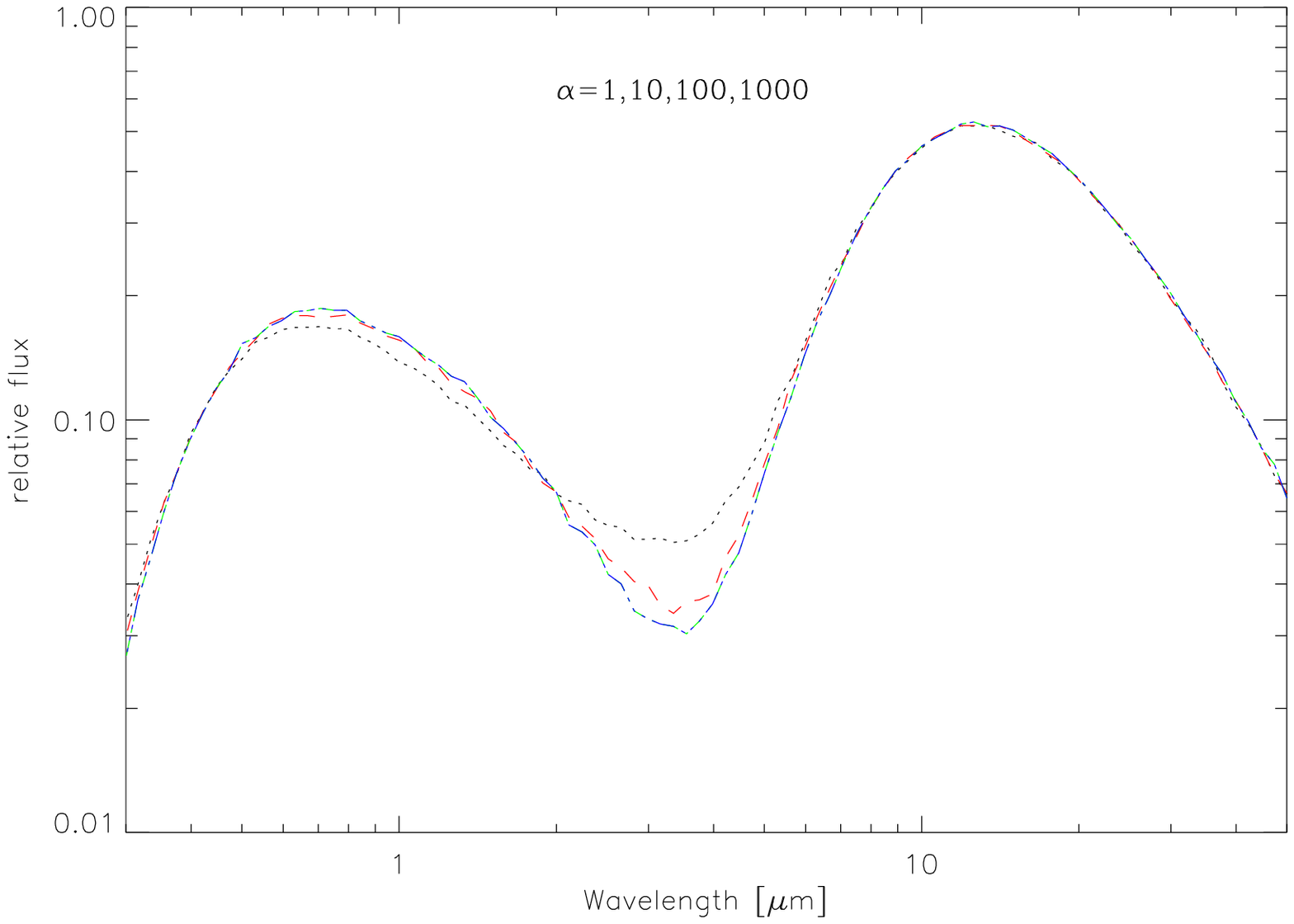}
\end{minipage}
\begin{minipage}[t]{8.5cm}
\includegraphics[width=8.5cm]{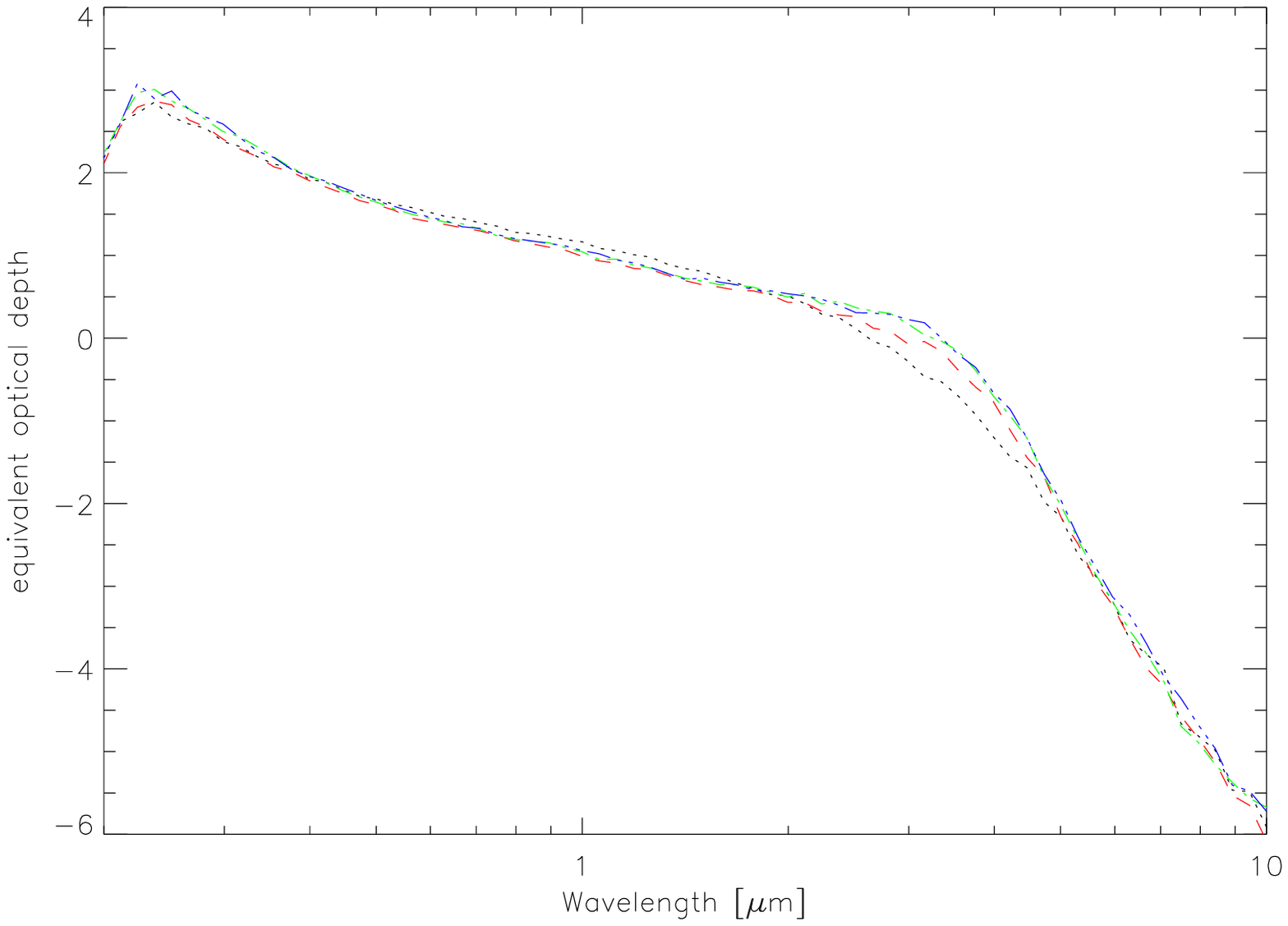}
\end{minipage}
\caption[]{Clumpy~{\sc i} models: dependence of SEDs (left panel) and
equivalent optical depths (right panel) on the clump to ICM density
enhancement factor, $\alpha$.  All models have the same $M_{\rm
d}$~=~5.6$\cdot$10$^{-4}M_{\odot}$ and $f$~=~0.05. Colour coding:
black (dotted) for $\alpha$~=~1, red (dashed) for $\alpha$~=~10, green
(dash-dot) for $\alpha$~=~100 and blue (dash-dot-dot) for
$\alpha$=1000. }
\label{fig:sedcealpha}
\end{center}
\end{figure*}

\begin{figure*}
\begin{center}
\begin{minipage}[t]{8.5cm}
\includegraphics[width=8.5cm]{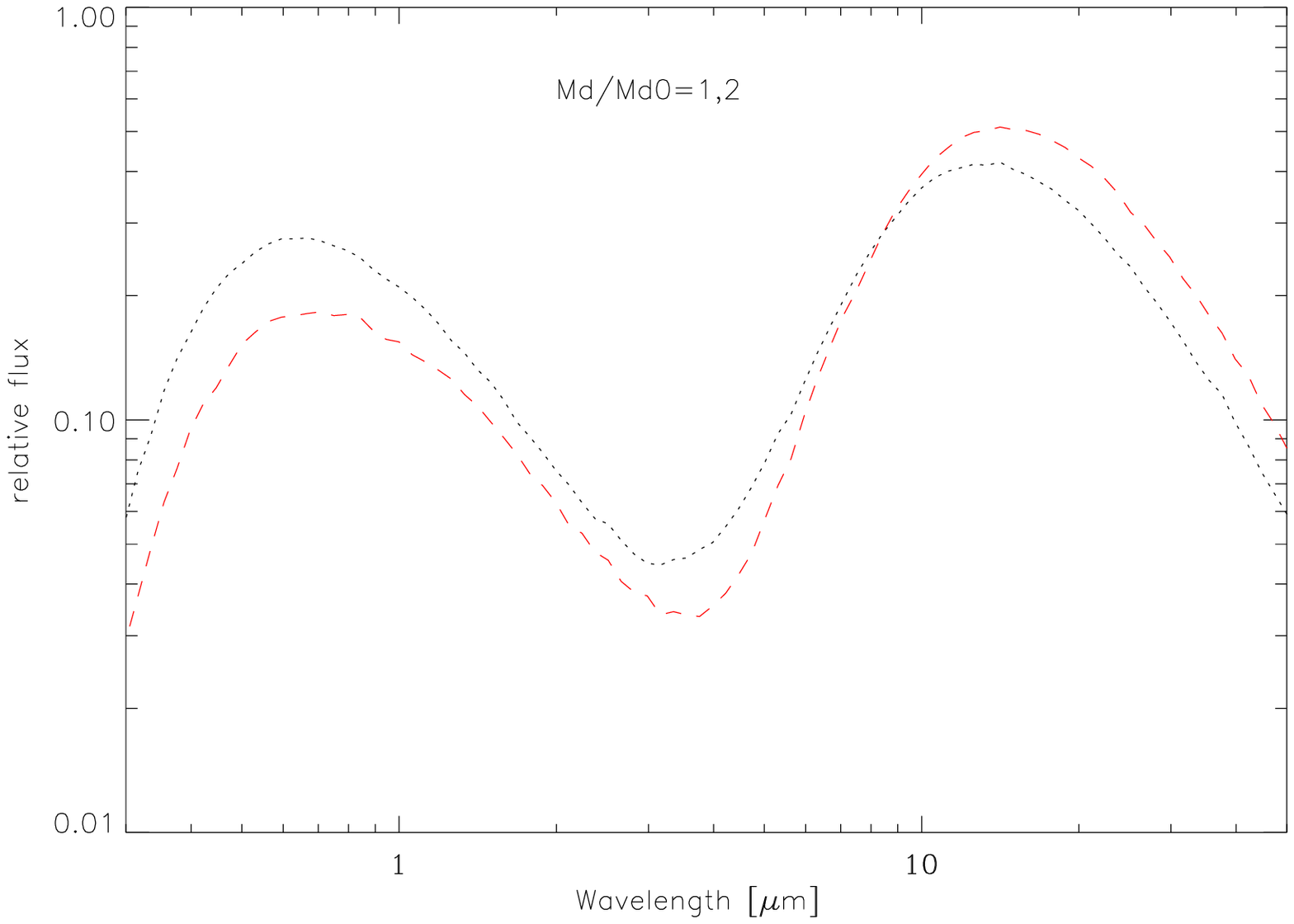}
\end{minipage}
\begin{minipage}[t]{8.5cm}
\includegraphics[width=8.5cm]{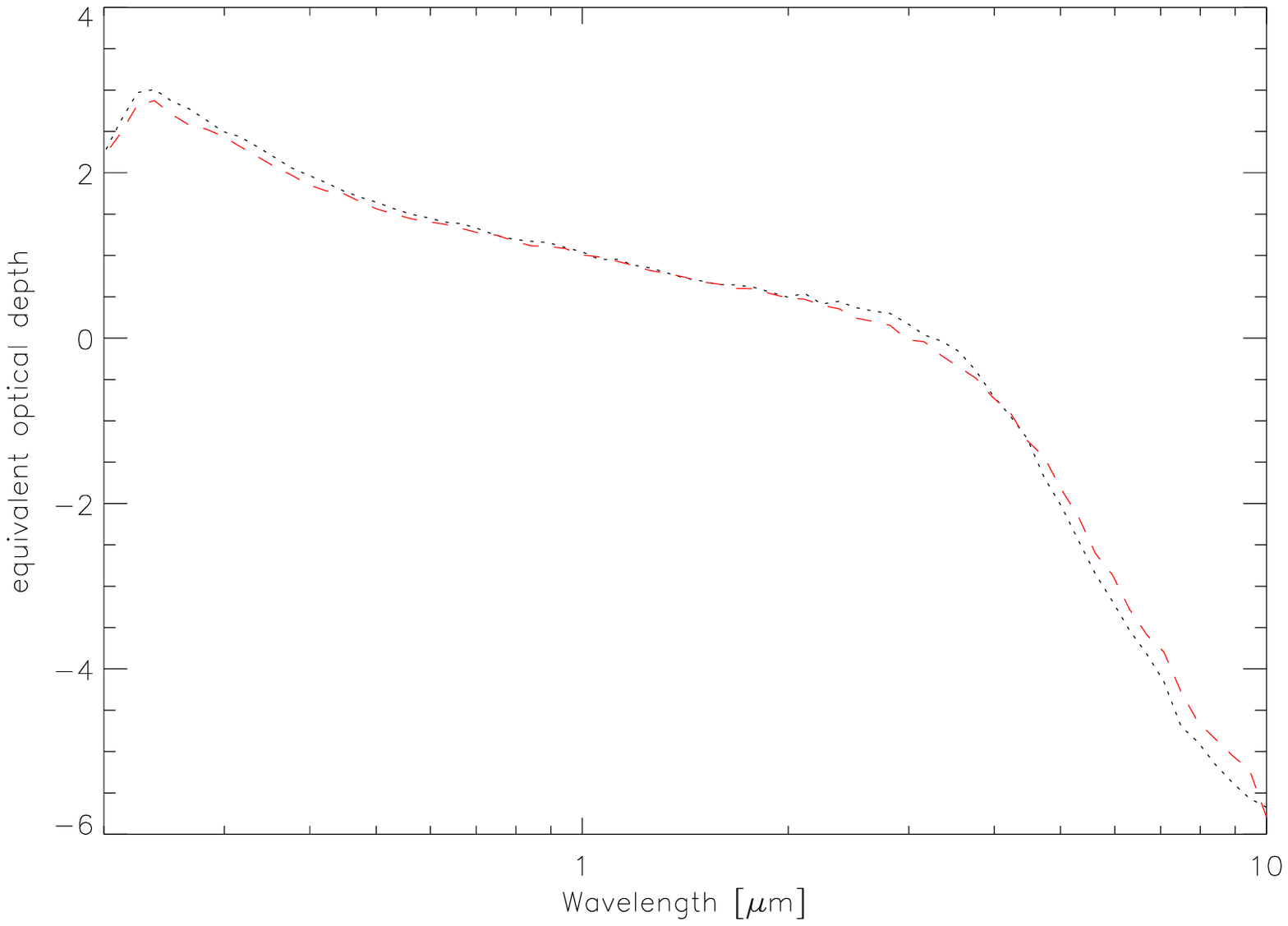}
\end{minipage}
\caption[]{Clumpy~{\sc i} models: dependence of SEDs (left panel) and
equivalent optical depths (right panel) on the total dust mass $M_{\rm
d}$.  All models have the same $\alpha$~=~100 and $f$~=~0.05. Colour
coding: black (dotted) for $M_{\rm d}$~=~5.6$\cdot$10$^{-4}M_{\odot}$,
red (dashed) for $M_{\rm d}$~=~11$\cdot$10$^{-4}M_{\odot}$.}
\label{fig:sedcemass}
\end{center}
\end{figure*}

\begin{figure*}
\begin{center}
\begin{minipage}[t]{8.5cm}
\includegraphics[width=8.5cm]{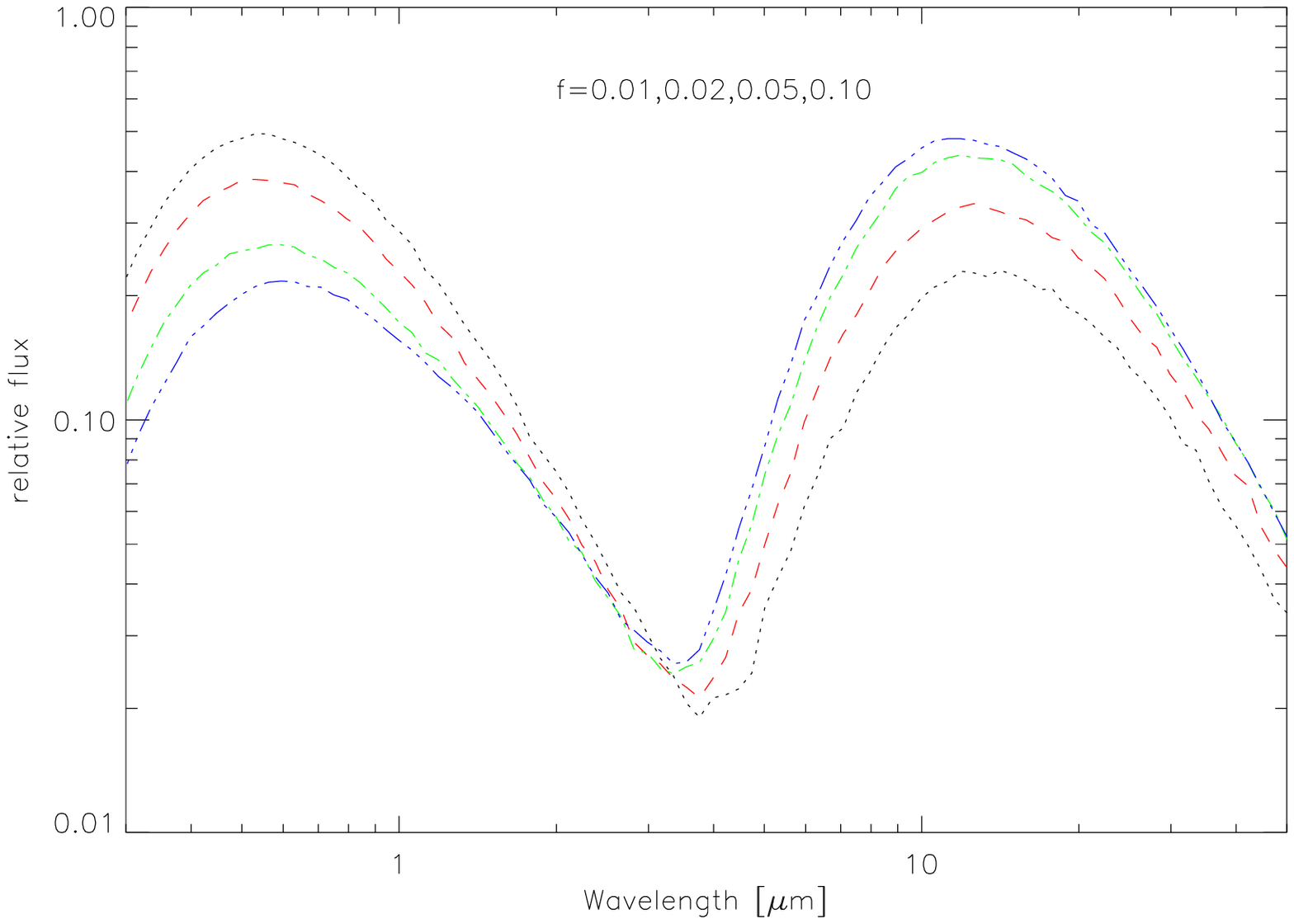}
\end{minipage}
\begin{minipage}[t]{8.5cm}
\includegraphics[width=8.5cm]{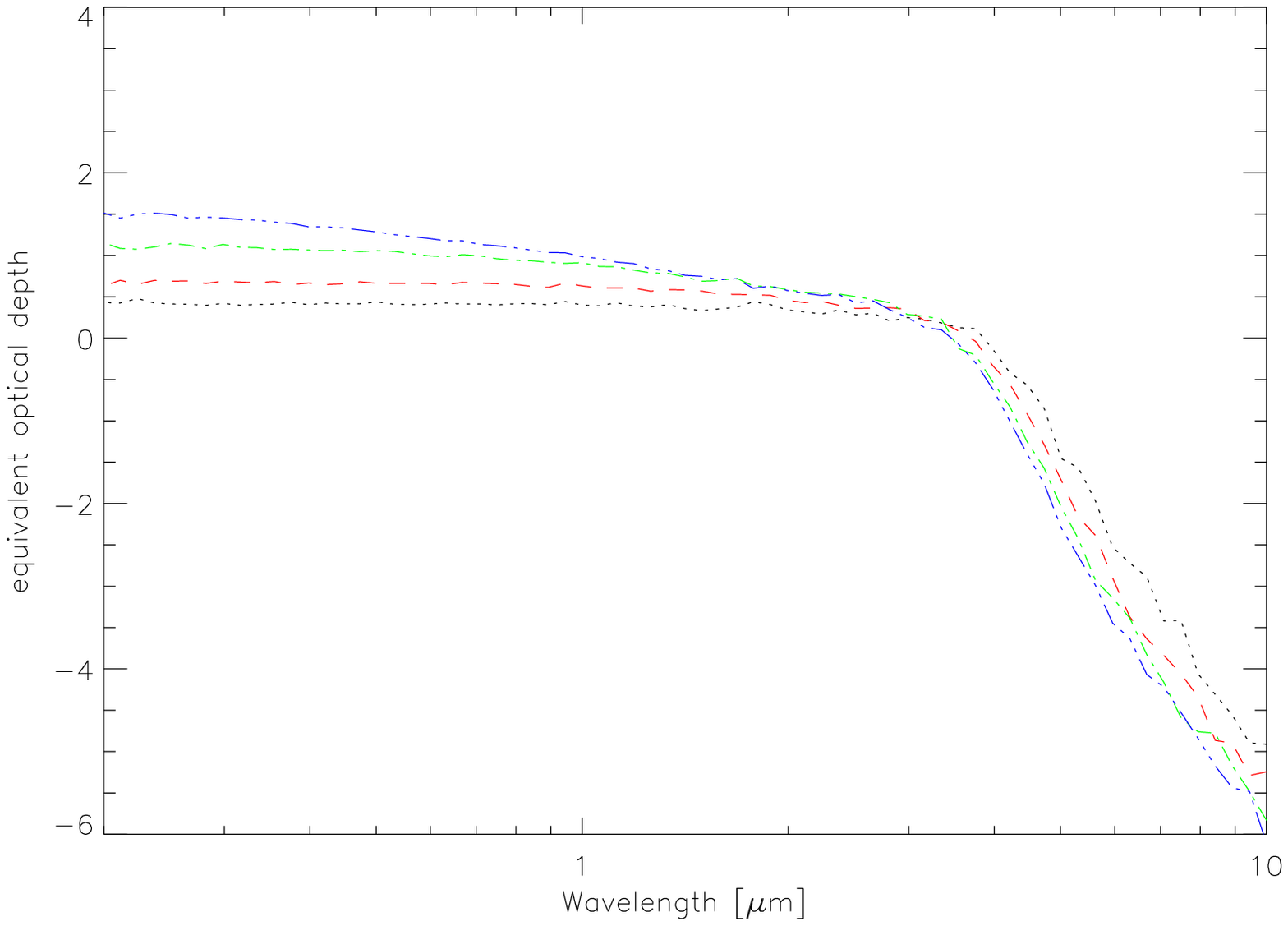}
\end{minipage}
\caption[]{Clumpy~{\sc ii} models: dependence of SEDs (left panel) and
equivalent optical depths (right panel) on the volume filling factor,
$f$, of the clumps. All models have the same $M_{\rm
d}$~=~5.6$\cdot$10$^{-4}M_{\odot}$ and $\alpha$~=~100.  Colour coding:
black (dotted) for $f$~=~0.01, red (dashed) for $f$~=~0.02, green
(dash-dot) for $f$~=~0.05 and blue (dash-dot-dot) for $f$=0.10.}
\label{fig:sedcenef}
\end{center}
\end{figure*}

\begin{figure*}
\begin{center}
\begin{minipage}[t]{8.5cm}
\includegraphics[width=8.5cm]{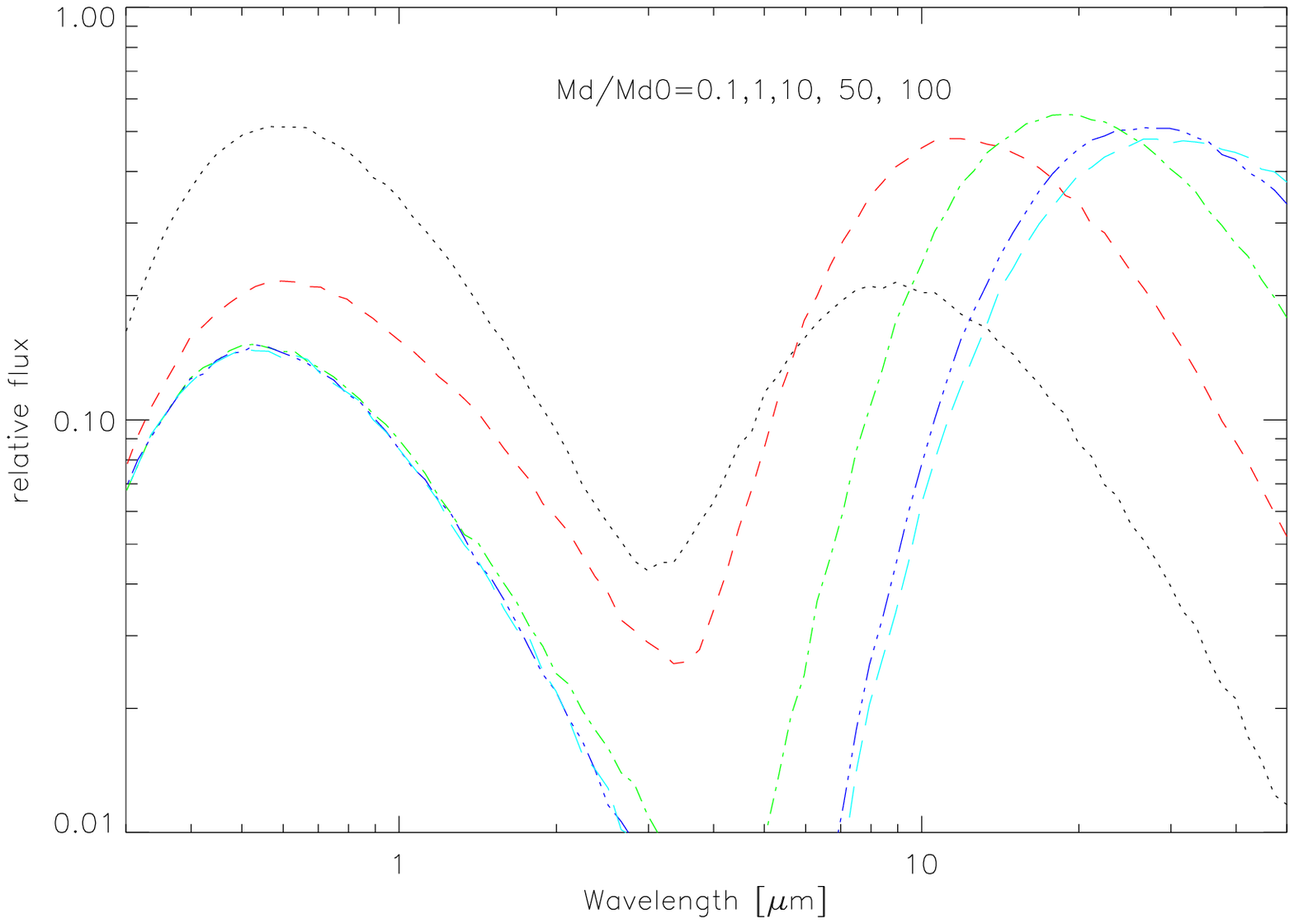}
\end{minipage}
\begin{minipage}[t]{8.5cm}
\includegraphics[width=8.5cm]{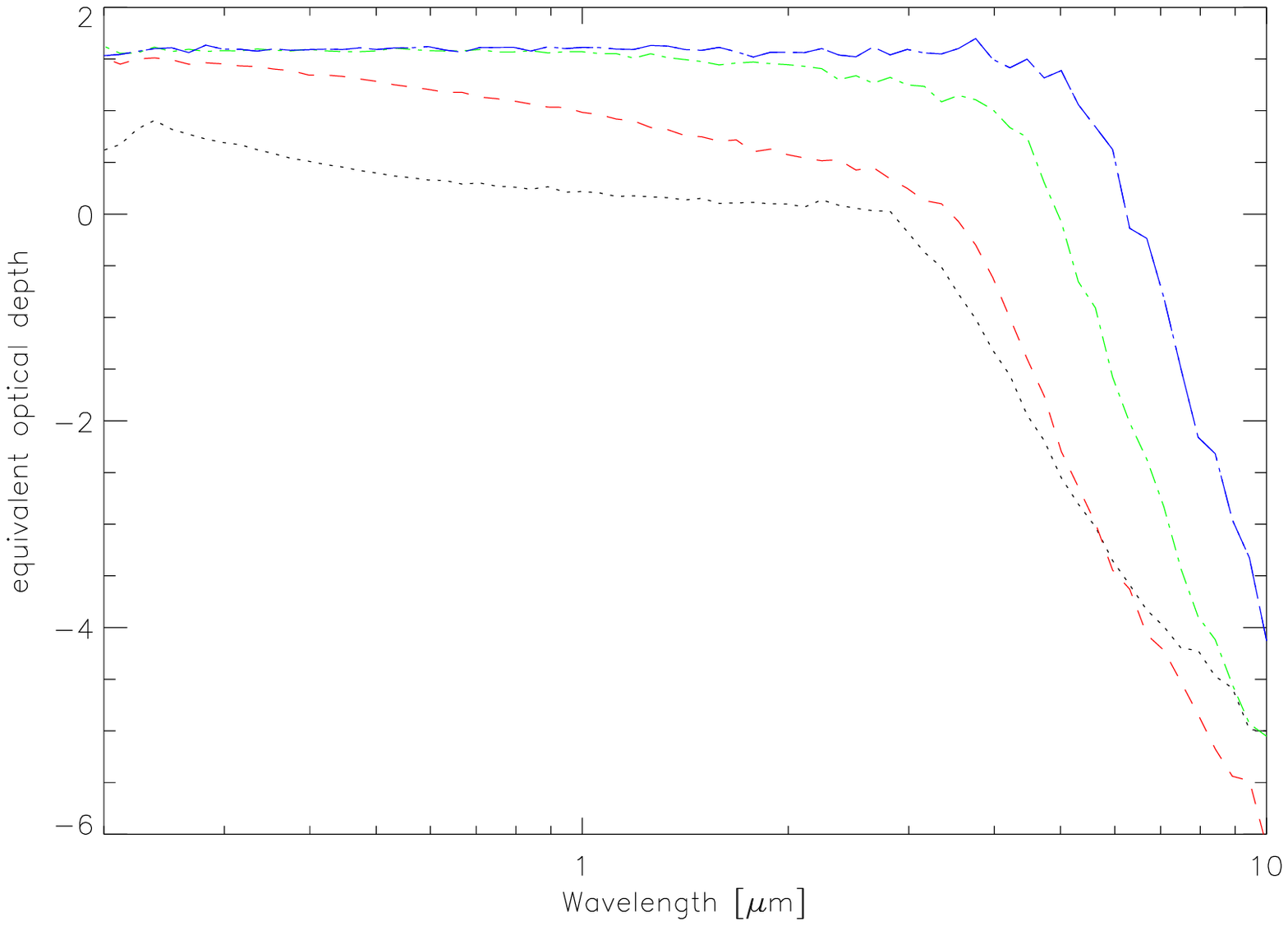}
\end{minipage}
\caption[]{Clumpy~{\sc ii} models: dependence of SEDs (left panel) and
equivalent optical depths (right panel) on the total dust mass $M_{\rm
d}$. All models have the same $\alpha$~=~100 and $f$~=~0.05. Colour
coding: red (dashed) for $M_{\rm d,0}$~=~3.5$\cdot$10$^{-4}M_{\odot}$,
black (dotted) for $M_{\rm d}$~=~0.1$\times M_{\rm d,0}$, green
(dash-dot) for $M_{\rm d}$~=~10$\times M_{\rm d,0}$, blue
(dash-dot-dot) for $M_{\rm d}$~=~50$\times M_{\rm d,0}$ and cyan (long
dash) for $M_{\rm d}$~=~100$\times M_{\rm d,0}$.}
\label{fig:sedcenemd}
\end{center}
\end{figure*}

\subsection{The 3D {\sc mocassin} code}

The 3D Monte Carlo photoionisation and dust RT code {\sc mocassin} was
described and benchmarked in \citep{Erc03,Erc05}.  We used
version 2.02.21 of the code in its dust-only RT mode, which allows a
multi-grid resolution approach in order to increase the spatial resolution
in the clumps. The radiation field is described by discrete monochromatic
packets of energy \citep{AL85}, whose trajectories through the
shell (mapped on multiple Cartesian grids) are characterised by
absorption, re-emission and scattering events, according to the local
medium opacities and emissivities. To ensure that energy conservation is
strictly enforced at each location, all energy packets are followed until
they escape the shell, contributing to the emergent SED.

The initial frequencies of the energy packets, emitted in this case 
by an extended diffuse source, are determined by sampling
the probability density functions (PDF's) obtained from the blackbody
distribution for the temperature of the source.  The total bolometric
luminosity, $L$, integrated over the diffuse source equals that of an
equivalent centrally-located point source and the energy carried by each
individual energy packet in unit time is simply $L/N_{\rm pack}$,
where $N_{\rm pack}$ is the total number of energy packets employed in the
simulation.

\subsection{The dust density distribution}
\label{sec:dd}

Our models consist of a spherical shell, with inner radius $R_{\rm
in}$, outer radius $R_{\rm out}~=~Y{\cdot}R_{\rm in}$ and a $\rho \propto r^{-n}$
density profile. 
Two basic dust density distributions are considered: one where the density
varies smoothly as $r^{-2}$ and the other with dense homogeneous clumps
embedded in a less dense ICM, with a density profile varying as $r^{-2}$.
The clumps have radius $\delta \cdot R_{out}$ and a volume filling factor,
$f$. The density contrast between the clumps and the ICM is defined by
$\alpha = \rho_{\rm c}(R_{\rm in})/\rho_{\sc icm}(R_{\rm in})$, where
$\rho_{\rm c}(R_{\rm in})$ and $\rho_{\sc icm}(R_{\rm in})$ are the clump
and ICM dust densities at the inner edge of the shell. The absolute dust
density in the clumps is therefore constant for all clumps, regardless of
their radial distance, meaning that the density contrast with the ICM is
larger at larger shell radii. Our choice to employ this model is based
upon the assumption that as early as a few hours after outburst,
post-shock ejecta become Rayleigh-Taylor unstable \citep{CK78,HW94},
forming a clumpy distribution. The clumps would then simply move outward
with the original imprint of density enhancement remaining intact. The
clump positions are assigned stochastically, with the probability of a
given grid cell being occupied by a clump being proportional to $r^{-2}$.
The total number of clumps in a given model is controlled by the volume
covering factor, $f$, and the clump size parameter, which is kept
constant at a value of $\delta$~=~$R_{\rm out}$/30.

The smooth models use 61$^3$ cells to describe the SN shell, although we
make use of the assumed symmetrical nature of the object and restrict our
simulations to only one eighth of the total volume, reflecting the packets
at the Cartesian planes \citep{Erc03b}.  Figure~\ref{fig:clumpy}
shows a 3D representation of the density distribution in the clumpy
models, consisting of cross-sections taken along the positive domains of
the orthogonal Cartesian planes. The clumpy models also use a {\it mother}
grid of 61$^3$ cells to describe the ICM, whilst each individual clump is
described by a subgrid of 5$^3$ cells which occupies the volume of a {\it
mother} grid cell. We experimented with higher resolution grids and 
subgrids and found that they had negligible consequences on our final 
results. 

For clumpy model configurations {\sc i} and {\sc ii} we first vary the
dominant clump parameters and show the effects of these variations on the
emergent SEDs and equivalent optical depths, defined in 
Section~\ref{sec:difil} below.

\subsection{The diffuse illuminating source and equivalent optical depths}
\label{sec:difil}

The illuminating source is adopted to have a total bolometric luminosity
$L~=~5.7\times10^5L_{\odot}$ at day 615 (from Figure~3 of W93) radiating
as a blackbody of $T_{\rm eff}$~=~7000~K. We experimented with lower
blackbody effective temperatures and came to the conclusion that the
observations could be best fitted by a value of T$_{\rm
  eff}$~=~7000~K, which may be interpreted as the superposition of the 
5000-5500~K optically thick `hot' component and
the H~{\sc i} bound-free continuum component identified in the W93 fits to
the SN~1987A observations (their Figure~2). We also note that for the same epoch (day 616)
\citet{SB90} and \citet{Bou91} found $L~=~6.6\times10^5L_{\odot}$ and 
$L~=~6.7\times10^5L_{\odot}$, respectively. However in this work, we adopt
the lower luminosity value which is consistent with and returns the better
fit to W93's data, as shown in Figure~\ref{fig:ltest}, where we
compare the SEDs obtained by identical smooth density models of day
615, but with different bolometric luminosity. The red-dotted line is for a
model with $L~=~5.7\times10^5L_{\odot}$ model and the green-dashed line is
for a model with $L~=~6.7\times10^5L_{\odot}$. The input parameters
for all models will be discussed in more details in the following sections.
  The diffuse source is distributed either
throughout the medium, including in the clumps (the Clumpy {\sc i} models,
in which the diffuse heating source intensity follows the dust density);
or located only in the ICM (the Clumpy {\sc ii} models), such that each
radiating cell $i$ emits a luminosity $L_i$ proportional to the local
interclump medium density.

In the case of a SN shell illuminated by a diffuse source of radiation
distributed within the shell, optical depths calculated along a line of
sight from the centre of the sphere have little observational relevance.  
To account for the fact that the photons emitted from different locations
within the shell will encounter very different opacities, we calculate the
{\it effective optical depth}, $\tau_{\rm eff}(\lambda)$, from the ratio
of the emerging unattenuated flux at $\lambda$ for the case where there is
no dust present, $F_{\lambda}^0$, to the emergent flux after attenuation
by dust in the shell, $F_{\lambda}$:
\begin{equation}
\tau_{\rm eff}(\lambda) = \ln\frac{F_{\lambda}^0}{F_{\lambda}}
\end{equation}
This method is completely independent of geometry and therefore does not
require averaging along a discrete number of lines of sight; furthermore
it automatically takes into account the effects of photons scattering off
the grains, which increases their path lengths through the shell, thus
increasing the probability of their being absorbed by dust before 
escaping.

\section{Results: the effect of clump parameters on the emergent SEDs}

\subsection{Clumpy {\sc i} models: diffuse heating source and dust 
located in both the clumps and in the ICM}

In this section we present results obtained from clumpy models where both
clumps and ICM radiate with luminosity $L_i$ proportional to the local
medium density. The effects of varying the volume filling factor of the
clumps, $f$, the clump-to-ICM density ratio, $\alpha$, and the dust mass,
$M_{\rm d}$ on the emerging SEDs and effective optical depths of the
models are illustrated in figures~\ref{fig:sedcef} to~\ref{fig:sedcemass}.  
The SEDs are plotted as dimensionless, distance- and
luminosity-independent spectral shapes, or relative flux,
$\lambda\,F_{\lambda}$/$\int{F_{\lambda}d\lambda}$, versus $\lambda$ in
microns..

Figure~\ref{fig:sedcef} shows that increasing the volume filling
factor $f$ of the clumps has dramatic effects on the effective optical
depth and therefore on the emergent SEDs, for models with
$\alpha$~=~100 and the same total dust mass $M_{\rm
d}$~=~5.6$\cdot$10$^{-4}M_{\odot}$, with the individual dust clumps
each having a mass of 10$^{-7}M_{\odot}$.  When $f$ = 0.05, each of
these clumps has a centre-to-edge radial optical depth in the R-band
of 1.22 for the amorphous carbon models. The same clumps containing
graphite grains \citep{DL84}, rather than amorphous carbon grains,
would have a centre-to-edge radial optical depth in the R-band of 0.46.

It is clear that an increase of the clump volume filling factor
causes a decrease in the effective optical depth of the models, with less
optical radiation being reprocessed at IR wavelengths. This somewhat
counter-intuitive effect is in fact a direct result of the luminosity
source being embedded in the clumps, so for constant $M_{\rm d}$ and
$\alpha$, an increase in $f$ implies a decrease in the clump dust density
and thus a reduction of the optical depth through each knot, facilitating
the escape of photons emitted by a given knot. This is opposite to the
behaviour of models which have a central source of radiation, or ones in
which only the ICM radiates, as described in the next section.

In figure~\ref{fig:sedcealpha} we show the SEDs (left panels) and the
equivalent optical depths (right panels) for models with a constant
$f$~=~0.05 and a total dust mass in all clumps of $M_{\rm
d}$~=~5.6$\cdot$10$^{-4}M_{\odot}$, for different values of $\alpha$.
Variations in $\alpha$ do not
appear to have large effects on the emerging SEDs, except perhaps for the
$\alpha$~=~1 model (red-dotted line), where there is no dust density enhancement in the
clumps with respect to the ICM. We note, however, that the ICM dust
density follows an $r^{-2}$ dependency, whereas the density of the clumps
stays constant with distance from the centre, implying that the clump to
ICM density ratio increases as the square of the distance from the centre.

The main point is that for a model where the intensity of the diffuse
heating source follows the dust density, dust grains cannot be hidden
in optically thick clumps. An increase in dust mass, e.g. a
factor of two, will change the optical and IR SED dramatically, as
shown in figure~\ref{fig:sedcemass}, where in the left- and right-hand
panels we plot the emergent SEDs and equivalent optical depths for
models where $f$ and $\alpha$ stay constant at values of 0.1 and 100,
respectively, and the total dust mass $M_{\rm d}$ varies from
$\sim$5$\cdot$10$^{-4}M_{\odot}$ to
$\sim$1$\cdot$10$^{-3}M_{\odot}$. Finally, it is clear that for these
models a degeneracy exists between $f$ and total dust mass $M_{\rm
d}$, which cannot be resolved solely by SED observations.

\subsection{Clumpy {\sc ii}: radiation from dust-free ICM and dust
confined to clumps only}

We now present the results obtained from clumpy models {\sc ii} having
exactly the same clump distribution as for {\sc i}, but this time
assuming that dust is only present in the clumps and that the luminosity
source is only embedded in the ICM. The parameter $\alpha$ tends to
infinity for these models, so we only investigate the effects of varying
$f$ and $M_{\rm d}$ on the emerging SEDs and equivalent optical depths.

The left and right panels of figure~\ref{fig:sedcenef} show the SEDs and
equivalent optical depths corresponding to models with constant total dust
mass $M_{\rm d}$~=~3.5$\cdot$10$^{-4}M_{\odot}$ and $f$~=~0.01 (black,
dotted), 0.02 (red, dashed), 0.05 (green, dash-dot) and 0.1 (blue,
dash-dot-dot). The left and right panels of figure~\ref{fig:sedcenemd}
show the SEDs and equivalent optical depths corresponding to models with
constant $f$~=~0.1 and $M_{\rm d}$ between $(0.35-3500)\times10^{-4}M_{\odot}$.
For the reference model, with 
$f$~=~0.1 and and $M_{\rm d}$~=~3.5$\cdot$10$^{-4}M_{\odot}$, each clump has
a mass of 3.5$\cdot$10$^{-8}M_{\odot}$ and a centre-to-edge radial optical
depth in the V-band of 1.33 and 0.41, for the amorphous carbon and graphite models,
respectively.

It is clear from figure~\ref{fig:sedcenef} that $f$ directly controls the
effective optical extinction, i.e. the amount of radiation intercepted by
the clumps and reprocessed at IR wavelengths. The behaviour of the SEDs
and equivalent optical depths here is the same as for a centrally located
source and opposite to that shown by the Clumpy {\sc i} models where the
heating radiation is emitted within both clumps and ICM.
 
In the current case of a dust-free ICM with the dust confined to clumps,
for optically thick clumps the equivalent optical depth in the visual
region is governed only by the covering factor of the clumps, which
determines the amount of radiation intercepted by the clumps. This would
imply that the actual dust mass in the clumps would be unconstrained by
the measured extinction in the optical and that virtually unlimited
amounts of dust could be hidden in optically thick clumps.  This is
apparent in figure~\ref{fig:sedcenemd}, where it can be seen that the
models with 10, 50 and 100$\times$~3.5$\cdot$10$^{-4}M_{\odot}$ show the
same SED between $\sim$0.3$\mu$m and $\sim$2$\mu$m. However the quantity
of dust in the clumps still has a significant effect on the grain
temperature distribution and therefore on the appearance of the SED at IR
wavelengths. The larger the dust mass for a given $f$, the cooler the
average grain temperature and therefore the greater the shift of the
reprocessed radiation peak to longer wavelengths.

So, for the Clumpy {\sc ii} case the total dust mass is not
unconstrained, even when the clumps are optically thick. If both optical
and IR observational data are available, it appears that the best
modelling strategy for optically thick clumps, when segregation between
the dust and the heating source is assumed, is (i) to start with a large
dust mass (enough for the clumps to be optically thick in the visual
region) and to vary $f$ until a fit is achieved for the optical part of
the SED; (ii) keeping $f$ fixed, to vary the dust mass until a fit to the
IR part of the SED is also achieved. If the dust mass is reduced to the
point where the optical SED starts changing again, this implies that the
clumps are becoming optically thin. This may be acceptable, or one may try
a lower covering factor and restart the fitting process.

\section{SN~1987A revisited}

\begin{table*}
\caption{Input parameters for SN~1987A dust shell models. The
parameters specified in the top section of the table are common to all
models. These include the total bolometric luminosity of the diffuse
illumination source, $L_*$, assumed to radiate as a blackbody of
temperature $T(BB)$, the inner shell radius, $R_{\rm in}$, and the
index $p$ of the power law describing the radial dependence of the
density distribution. The carbon dust is either 100\% amorphous carbon
\citep[optical constants from][]{Han88} or 100\% graphite
\citep[optical constants from][]{DL84}, with a standard MRN grain size
distribution truncated at the upper end ($a_{min}$~=~0.005~$\mu$m;
$a_{max}$~=~0.05~$\mu$m). Both smooth and clumpy models with a diffuse
illumination source are considered.  For the smooth models the dust
number density decreases as $r^{-2}$. The Clumpy~{\sc i} models have
radius-dependent clump-interclump dust number density enhancement
factors, $\alpha$, corresponding to the enhancement at the inner
radius of the shell. Clumps are randomly distributed within the shell,
according to a probability density function proportional to the square
of the inverse of the radial distance.  The clump volume filling
factors, $f$, are also given for each model, as well as the total dust
masses, $M_{\rm d}$, for both smooth and clumpy models with amorphous
carbon or graphite grain and the outer-to-inner shell radius ratio, $Y =
R_{\rm out}/R_{\rm in}$.
The R-band extinctions, $A_R$, are also listed 
for each model, as well as the radial centre-to-edge extinction within an individual 
clump, $A_R$(clump).
} 
\label{tab:bestfit}
\begin{center}
\setlength\tabcolsep{5pt}
\begin{tabular}{lccccccc}
\hline\noalign{\smallskip}
                                & \multicolumn{3}{c}{Day 615}       & 
~~~~~~~~~         &  \multicolumn{3}{c}{Day 775}                \\
\hline\noalign{\smallskip}
$L_* [10^{5}L_{\odot}]$       &        &    5.7         &       & ~~~~~~~~~            &        &  1.6           &                    \\
$T(BB) [K]$                   &        &    7000        &       & ~~~~~~~~~            &        &  7000          &                    \\
$R_{in} [10^{15}$cm$]$        &        &     5          &       & ~~~~~~~~~            &        &   5.7          &                    \\
$p$                           &        &     2          &       & ~~~~~~~~~            &        &   2            &                    \\
\hline\noalign{\smallskip}
          Amorphous Carbon & Smooth & Clumpy~{\sc i} &  Clumpy~{\sc ii}  & ~~~~~~~~~ & Smooth & Clumpy~{\sc i} &  Clumpy~{\sc ii}\\
\hline\noalign{\smallskip}
 $\alpha$                       &  --    &     100        &   $\infty$        & ~~~~~~~~~ &  --    &      100       &    $\infty$      \\
 $f$                            &  --    &     0.05       &     0.1           & ~~~~~~~~~ &  --    &      0.03      &      0.3         \\
$R_{\rm out}/R_{\rm in}$      &    7    &     5          &    5   & ~~~~~~~~~ &   7     &   5     &        5            \\
 $M_{\rm d} [M_{\odot}]$ &  2.0$\cdot10^{-4}$ & 2.0$\cdot10^{-4}$  &  2.2$\cdot10^{-4}$ & ~~~~~~~~~ & 3.0$\cdot10^{-4}$ & 2.0$\cdot10^{-4}$  &  4.2$\cdot10^{-4}$        \\
 $A_R$ &  1.6  &     1.4     &      1.5     & ~~~~~~~~~ &   1.8   &      2.0   &    1.6  \\
 $A_R$(clump) &  --  &    2.2    &    1.2     & ~~~~~~~~~ &   --   &     2.8   &    0.60  \\
\hline\noalign{\smallskip}
 Graphite & Smooth & Clumpy~{\sc i} &  Clumpy~{\sc ii}  & ~~~~~~~~~ & Smooth & Clumpy~{\sc i} &  Clumpy~{\sc ii}\\
\hline\noalign{\smallskip}
 $\alpha$                       &  --    &     100        &   $\infty$        & ~~~~~~~~~ &  --    &      100       &    $\infty$      \\
 $f$                            &  --    &     0.05       &     0.1           & ~~~~~~~~~ &  --    &      0.03      &      0.3         \\
$R_{\rm out}/R_{\rm in}$      &    7    &     5          &    5   & ~~~~~~~~~ &   4     &   5     &        4            \\
 $M_{\rm d} [M_{\odot}]$ &  4.2$\cdot10^{-4}$ & 6.5$\cdot10^{-4}$ & 6.6$\cdot10^{-4}$ & ~~~~~~~~~ &  4.5$\cdot10^{-4}$ &  6.5$\cdot10^{-4}$  &  7.5$\cdot10^{-4}$       \\
 $A_R$ &    1.3   &   1.5  &      1.4    & ~~~~~~~~~ &  2.4   &    2.1     &   1.5        \\
 $A_R$(clump) &  --  &    2.5    &   1.2     & ~~~~~~~~~ &   --   &    3.3   &  0.89  \\
\hline
\end{tabular}
\end{center}
\end{table*}

Taking into account the results presented above, we now derive revised
dust mass estimates for SN~1987A. Our analysis is based on model fits
to the 0.3--30\,$\mu$m spectrophotometry from days 615 and 775
published in figures 1 and 2 of W93, which were digitised using the
{\it Dexter} tool provided by the NASA ADS, then dereddened using the
\citet{Mat90} extinction law, scaled to fit the extinction measured
toward SN 1987A in \citet{Scu96}.  Our models are also required to
match the optical extinction $A_R$ of the SN, which was first shown to
increase with time by \citet{Luc91}.  However, as discussed in
\citet{Sug06}, the method adopted by \citet{Luc91} to measure the dust
extinction from the broad-band light curves is likely to have
significantly underestimated the $V$ and $R$-band values.  



\begin{figure}
\begin{center}
\begin{minipage}[t]{8.5cm}
\includegraphics[width=8.5cm]{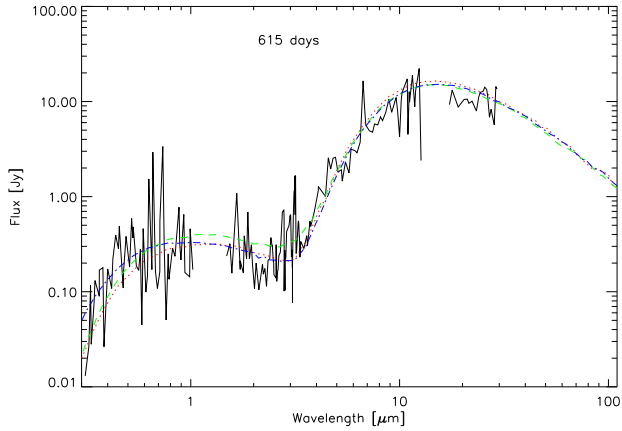}
\end{minipage}
\caption[]{Amorphous carbon model fits to the day 615 spectrophotometry 
published by W93 (black line). All the models 
used an MRN size distribution truncated to maximum grain radii 
of 0.05~$\mu$m. The red dotted line corresponds to a smooth density 
distribution model, the dashed green line to our Clumpy~{\sc i} model and 
the blue dot-dash line to our Clumpy~{\sc ii} models. Model parameters 
are summarised in Table~\ref{tab:bestfit}.}
\label{fig:sn87a615}
\end{center}
\end{figure}

\begin{figure}
\begin{center}
\begin{minipage}[t]{8.5cm}
\includegraphics[width=8.5cm]{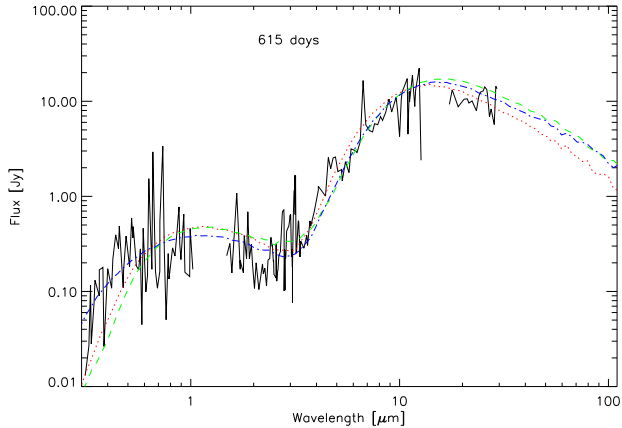}
\end{minipage}
\caption[]{Same as Fig.\ \ref{fig:sn87a615} but for graphite model fits.}
\label{fig:sn87a615gr}
\end{center}
\end{figure}

\begin{figure}
\begin{center}
\begin{minipage}[t]{8.5cm}
\includegraphics[width=8.5cm]{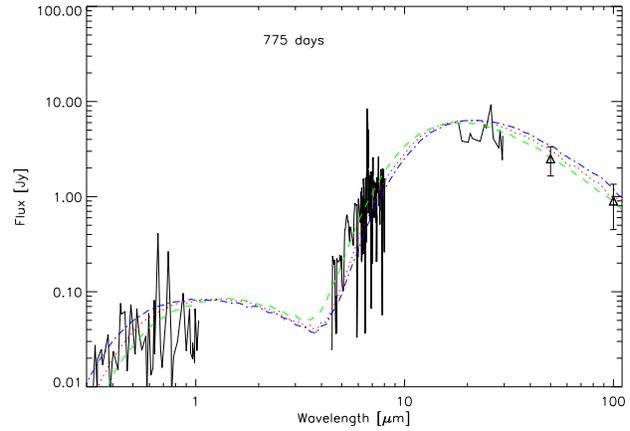}
\end{minipage}
\caption[]{Amorphous carbon model fits to the day 775 0.3--30-$\mu$m
  spectrophotometry published by W93 (black line) and the day 793
  50- and 100-$\mu$m photometry of \citet{Har89}
  (triangles). All models used an MRN size distribution truncated to a
  maximum grain radius of 0.05~$\mu$m. The red dotted line corresponds
  to a smooth density distribution model, the dashed green line to our
  Clumpy~{\sc i} model and the blue dot-dash line to our Clumpy~{\sc
    ii} models. Model parameters are summarised in
  Table~\ref{tab:bestfit}.  }
\label{fig:sn87a775}
\end{center}
\end{figure}

\begin{figure}
\begin{center}
\begin{minipage}[t]{8.5cm}
\includegraphics[width=8.5cm]{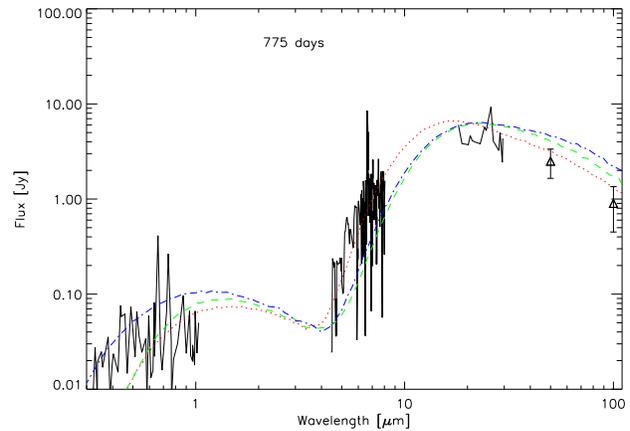}
\end{minipage}
\caption[]{Same as Fig.\
  \ref{fig:sn87a775} but for graphite models.}
\label{fig:sn87a775gr}
\end{center}
\end{figure}

Our smooth density distribution and our Clumpy~{\sc ii} models for day 615
(red dotted and blue dash dot lines in Figure~\ref{fig:sn87a615} for
amorphous carbon and Figure~\ref{fig:sn87a615gr} for graphite) both fit
the IR and optical SED of SN~1987A quite well -- the
smooth density distribution graphite model (Figure 8) provides a somewhat
better fit to the 20-$\mu$m spectrophotometry than the clumped models, but
at the expense of under-predicting the flux level in the short wavelength
optical region. The effective extinction predicted by our Clumpy~{\sc ii} models 
for day 615 is A(U)~=~1.69, A(B)~=~1.66, A(V)~=~1.61, A(R)~=~1.54 and A(I)~=~1.48, for amorphous 
carbon grains and A(U)~=~1.64, A(B)~=~1.54, A(V)~=~1.44, A(R)~=~1.38 and A(I)~=~1.31, 
for graphite grains. 
Although our absolute values are larger than those reported by
\citet{Luc91} -- see discussion in \citet{Sug06} about the extinction
being underestimated by the method
adopted by \citet{Luc91} --, the slope of the computed extinctions is in
agreement with the results of \citet{Luc91}, who noted that the empirical extinctions were rather flat in the optical. 
Our day 615 Clumpy~{\sc i}
models, whilst providing satisfactory fits to the IR SED, provides a
poorer match in the optical region. For the day 775 observations
(Figure~\ref{fig:sn87a775} for amorphous carbon and
Figure~\ref{fig:sn87a775gr} for graphite), the smooth and Clumpy~{\sc i}
models produce too much extinction at the shorter optical wavelengths,
particularly for the graphite case.

All inputs for our models are listed in Table~1, where we also
summarise our dust mass estimates, R-band model extinctions and 
the R-band radial centre-to-edge extinction within an individual clump.

\begin{figure}
\begin{center}
\begin{minipage}[t]{8.5cm}
\includegraphics[width=8.5cm]{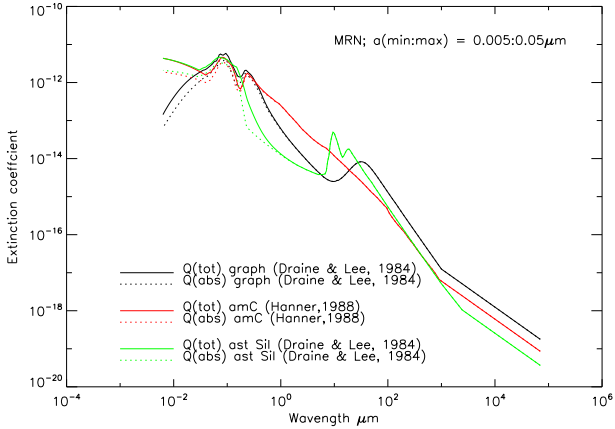}
\end{minipage}
\caption[]{Absorption and total extinction coefficients ($Q$) of
\citet{DL84} graphite compared to \citet{Han88} amorphous carbon and
to \citet{DL84} astronomical silicate. The coefficients were computed
using Mie scattering routines using an MRN size distribution with
minimum and maximum grain radii of 0.005 and 0.05 $\mu$m, identical to
that used in all models presented in this paper.}
\label{fig:amCvGraphvSilMRN}
\end{center}
\end{figure}

The day 615 smooth density distribution amorphous carbon model requires a
total dust mass of 2.0$\cdot$10$^{-4}$M$_{\odot}$, while the amorphous
carbon dust masses derived from our Clumpy~{\sc i} and Clumpy~{\sc ii}
models are the same within the uncertainties,
2.0$\cdot$10$^{-4}$M$_{\odot}$ and 2.2$\cdot$10$^{-4}$M$_{\odot}$,
respectively. Our graphite grain models require approximately 3 times more
mass to fit the SED of SN~1987A than do our amorphous carbon models, due
to the significantly lower optical to mid-IR absorption coefficients of
\citet{DL84} graphite compared to Hanner (1988) amorphous carbon
(see Figure~\ref{fig:amCvGraphvSilMRN}). Our smooth density distribution
graphite model requires a dust mass of 4.2$\cdot$10$^{-4}$M$_{\odot}$, a
factor of 10 higher than estimated by W93 for day 615 using their simple
large-grain analytic model. Our Clumpy~{\sc i} and {\sc ii} graphite
models both require $\sim$6.5$\cdot$10$^{-4}$M$_{\odot}$, factor of two larger
than derived by W93 for day 615 using an analytical clumpy model with
graphite absorption coefficients.

For day 775, our smooth density distribution amorphous carbon model
requires a dust mass of 3.0$\cdot$10$^{-4}$M$_{\odot}$, while our
Clumpy~{\sc i} and Clumpy~{\sc ii} models require
2.0$\cdot$10$^{-4}$M$_{\odot}$ and 4.2$\cdot$10$^{-4}$M$_{\odot}$,
respectively. Our smooth, Clumpy~{\sc i} and Clumpy~{\sc ii} graphite
models (Fig.\ \ref{fig:sn87a775gr}) require masses of
4.5$\cdot$10$^{-4}$M$_{\odot}$, 6.5$\cdot$10$^{-4}$M$_{\odot}$ and
7.5$\cdot$10$^{-4}$M$_{\odot}$ respectively to fit the day 775 SN~1987A
data. The day 775 smooth
density distribution amorphous carbon and graphite models both provide
better fits to the observed SEDs landwards of 20~$\mu$m than do the clumpy
models, but both of the smooth density distribution models fail to match
the observed flux level in the short wavelength optical region.
The IR spectrum seems to be too sharply peaked at short wavelengths to
be fitted with simple models. The residuals are also broader than
typical atomic line widths, which could be interpreted as evidence for 
solid-state features in the dust --possibly stochastically heated--, 
as discussed Bouchet et al., 2004.
Our smooth density distribution graphite mass estimate is 
7 times larger than W93's analytic large-grain estimate for this epoch,
while our Clumpy~{\sc i} and Clumpy~{\sc ii} graphite grain mass estimates 
are factors of 1.3--1.5 larger than W93's analytic clumpy model estimate of
5.0$\cdot$10$^{-4}$M$_{\odot}$ for graphite grains. Dwek et al. (1992)
also estimated a dust mass of 5$\cdot$10$^{-4}$M$_{\odot}$ for the
ejecta of SN~1987A, based on fitting a 150~K modified blackbody to day
1144 IR photometric data. Although their estimate is somewhat
uncertain, due to the small number of photometric points in the mid-IR
region, it does indicate that the mass of dust in the ejecta of
SN~1987A did not increase significantly after day 775. Furthermore,
 \citet{Bou04} presented mid-IR imaging of SN~1987A and reported, in
 particular, the detection of the SN ejecta in the N-band. They
 concluded that their measurements imply a dust temperature in the 
90--100~K range, and a mass range of
$0.1-2\cdot10^{-3}~M_{\odot}$, again ruling out a significant
variation in dust mass between day 775 and day 6067. 
Our own dust mass
estimates in Table~\ref{tab:bestfit} imply that for the graphite models
the total dust mass hardly increased between days 615 and 775.
Finally, Bouchet et al. (2004) detected mid-IR silicate emission
from the circumstellar ring that is believed to have been produced by mass 
loss by the progenitor star of SN~1987A. This ring is responsible for
most of the present-day mid-IR emission, but Bouchet et al. (2004) 
showed that this contribution started after day 4200 only, and therefore does
not affect the present findings.   

We also made estimates for the circumstellar extinction in the
envelope of SN~1987A at late epochs in order to constrain our
models. For this we used the methods described in greater detail by
\citet{Sug06} for the case of SN~2003gd. Using the photometric data of
\citet{Ham90} for days 1-813 and \citet{Cal93} thereafter, we
estimated the circumstellar R-band extinction of SN~1987A at days 615
and 775 by comparing the observed level of the light curve at those
epochs to that expected for a light curve powered purely by the decay
of $^{56}$Co that had been normalised to the photometric data obtained
earlier than day 400.  This yielded $A_R$'s of 1.7 and 2.4 magnitudes
at days 615 and 775, respectively. A somewhat more conservative
estimate of the circumstellar extinction can be obtained if we assume
that the intrinsic light curve has contributions from other sources of
decay, such as $^{57}$Co, and also allow for the decreasing opacity of
the envelope to the gamma rays that power the light curve. This
yielded $A_R$'s of 1.2 and 1.6 magnitudes at days 615 and 775,
respectively. Corresponding estimates for the R-band circumstellar
extinction produced by the models were obtained by comparing the
R-band flux levels obtained with and without dust. Inspection of
Table~1 (final row for each model) shows that all of our day 615
models satisfy the $A_R$ = 1.2--1.7 observational constraint. For the
case of our day 775 models the predicted $A_R$ range is 1.5--2.4,
which can be compared to our estimated observational constraint of
$A_R$ = 1.6--2.4.

For Clumpy~{\sc i} amorphous carbon or graphite models, the derived day 615 and day
775 dust masses are the same, while the inner radius for the day 775
models is 5.7/5.0 times larger than for day 615. Yet the day 775
models yield higher $A_R$'s than the day 615 models.  The equivalent
optical depth increase from day 615 to 775 is due to the fact that for
the volume filling factor to decrease from 5 to 3\% whilst keeping the
same dust mass , the dust number density in each individual clump must
increase, hence making it harder for radiation produced in the
internal regions of the clumps to escape. The decrease of the clump
covering factor from day 615 to 775 is physically justified under the
assumption that no clumps and no extra dust condensed between day 615
to day 775. If this were true then the expansion of the inner radius
from 5$\times$10$^{15}$cm to 5.7$\times$10$^{15}$cm would imply a
decrease of the volume covering factor from $f$\,=\,0.05 to
$f$\,=\,0.034. However, best fit of the SED implies $f$ = 0.03, causing
therefore the increase of $A_R$ from 1.4 to 2.0 reported in Table~1.
 
The behaviour of Clumpy~{\sc ii} models is somewhat different; the
volume filling factor here has to increase from 0.1 to 0.3 in order to
fit the decrease of the SED in the optical from day 615 to day
775. As discussed in Section~3.2, once the clumps become optically
thick, the only factor affecting the optical SED is $f$, with the main
effect of varying the dust mass being on the temperature of the grains
and therefore on the IR emission. In this model the total dust mass
also increases between the two epochs, implying that as new grains
form new clumps also form. An alternative interpretation is that
clumps form early on by Rayleigh-Taylor instabilities in the explosion,
and remain imprinted in the ejecta. The increase in dust mass, might
therefore be due to additional accretion.

\begin{figure}
\begin{center}
\begin{minipage}[t]{8.5cm}
\includegraphics[width=8.5cm]{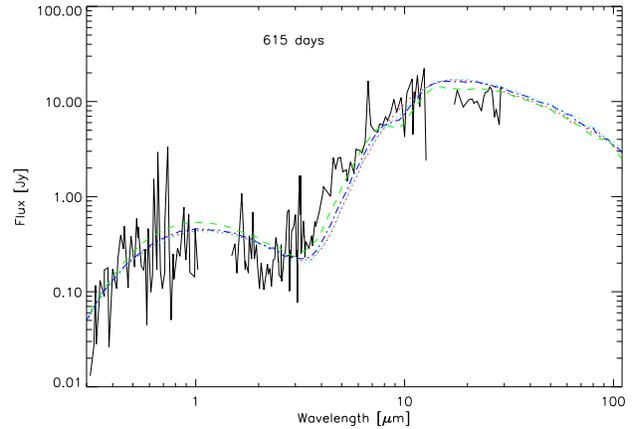}
\end{minipage}
\caption[]{Alternative Clumpy~{\sc ii} model fits to the day 615
optical-IR spectrophotometry published by Wooden et al. (1993) --black
line-- . All models used an MRN size distribution truncated to maximum
grain radii of 0.05$m$m. The red dotted line corresponds to a graphite
\citep{DL84} model. The dashed green line corresponds to a model with
a composition of 50\%:50\% astronomical silicates : graphite by mass
\citep{DL84}, the blue dash-dot line corresponds to 25\%:75\%
astronomical silicates : graphite, and the cyan dotted line
corresponds to 15\%:85\% astronomical silicates : graphite \citep{DL84}. The mixed composition models shown all correspond to a
dust mass of 1.3$\cdot$10$^{-3}$~M$_{\odot}$, twice as large as for
the Clumpy~{\sc ii} day 615 graphite model that is plotted.
}
\label{fig:sp}
\end{center}
\end{figure}

\begin{figure}
\begin{center}
\begin{minipage}[t]{8.5cm}
\includegraphics[width=8.5cm]{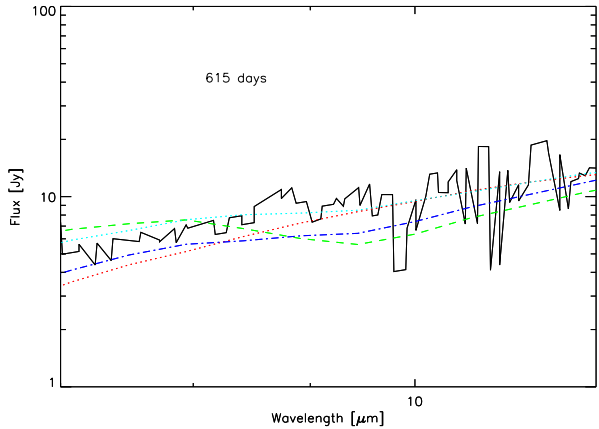}
\end{minipage}
\caption[]{Alternative Clumpy~{\sc ii} models (as in Figure~\ref{fig:sp})
  superimposed to the 7--12~$m$m region of the observed day 615
  spectrum from Figure~1 of W93. The red dotted line corresponds to a graphite
\citep{DL84} model. The dashed green line corresponds to a model with
a composition of 50\%:50\% astronomical silicates : graphite by mass
\citep{DL84}, the blue dash-dot line corresponds to 25\%:75\%
astronomical silicates : graphite, and the cyan dotted line
corresponds to 15\%:85\% astronomical silicates : graphite \citep{DL84}. } 
\label{fig:spzoom}
\end{center}
\end{figure}

Our dust mass estimates are dependent on the assumed dust
chemistry. As shown in Figure~\ref{fig:amCvGraphvSilMRN}, amorphous
carbon grains with Hanner (1998) optical constants have higher
absorption coefficients per unit mass in the optical and near-IR than
either astronomical graphite or silicate grains with \citet{DL84}
optical constants. The amorphous carbon models therefore require the
least mass to fit the observed SEDs of SN~1987A. Graphite models
require between 2 and 3 times as much mass to provide a
fit. Silicate-only models would require even larger dust masses, due
to their even lower optical and near-IR absorption coefficients;
however pure silicate models are ruled out for SN~1987A by the lack of
detectable 10- and 18-$\mu$m silicate emission or absorption features
in any of the mid-IR spectra of SN~1987A (W93; Roche et al. 1993;
Bouchet \& Danziger 1993).

After day 530, the emission-line strengths of [Si~{\sc i}] 1.65$\mu$m,
[Mg~{\sc i}] 4571\AA, and [O~{\sc i}] 6300\AA\ decreased significantly
\citep{Luc91,BD93}, consistent with the formation of {\em some}
fraction of silicate dust. We therefore tried a number of mixtures of
astronomical silicates and graphite.  Figure~\ref{fig:sp} shows fits
to the day 615 W93 data obtained with our Clumpy~{\sc ii} model using
50\%~:~50\% (green dash-dot line), 25\%~:~75\% (blue dash-dot line)
and 15\%~:~85\% (cyan dash-dot-dot line) astronomical
silicate~:~graphite mixtures, all with masses of
1.3$\cdot$10$^{-3}$M$_{\odot}$ (red dotted line), i.e. twice as large
as for the Clumpy~{\sc ii} graphite model. The 50\%~:~50\%
silicate:graphite and 25\%~:~75\% mixtures produce too strong 10- and
18-$\mu$m silicate absorption features compared to the observational
limits. However, the 15\%~:~85\% silicate:graphite model produces a
smooth enough mid-IR spectrum to be consistent with the observations.

Due to the high optical absorption coefficients of amorphous carbon
grains, the Clumpy~{\sc ii} day 615 amorphous carbon mass estimate of
2.2$\cdot$10$^{-4}$~M$_{\odot}$ (Table~\ref{tab:bestfit}) is likely to
represent a lower limit to the total mass of dust that condensed in the
ejecta of SN~1987A, while the mixed silicate:graphite mass
of 1.3$\cdot$10$^{-3}$~M$_{\odot}$, discussed in the previous 
paragraph, probably represents close to
an upper limit. Iron grains have optical absorption coefficients in the
optical that are intermediate between those of astronomical graphite and
silicate, while lacking the strong infrared features that eliminate
silicate-only models. However, direct observational evidence for iron
grains in circumstellar or interstellar environments is still lacking.


\begin{figure}
\begin{center}
\begin{minipage}[t]{8.5cm}
\includegraphics[width=8.5cm]{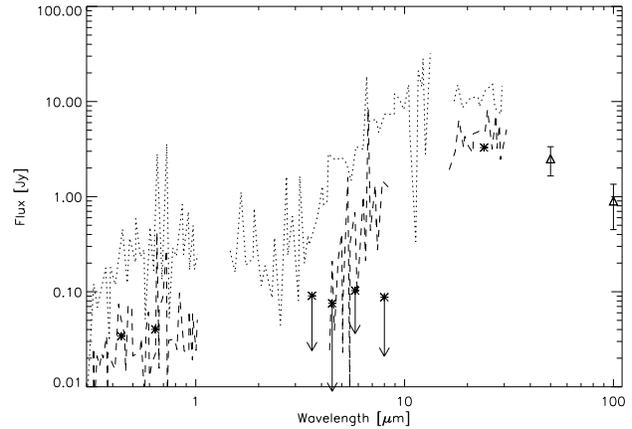}
\end{minipage}
\caption[]{The observed SED for day 678 of SN~2003gd (asterisks; Sugerman 
et al. 2006, and references therein) compared to those of SN~1987A on day 
615 (dotted line) and day 775 (dashed line). The optical to mid-IR 
spectrophotometry of SN~1987A is from W93
and the 50- and 100-$\mu$m photometric points at day 793
(triangles) are from \citet{Har89}.}
\label{fig:comp}
\end{center}
\end{figure}

Figure~\ref{fig:comp} shows a comparison between the day 678 optical-IR
spectrum of SN~2003gd \citep{Sug06} and the day 615 and day 775 spectra of
SN~1987A (W93), after scaling to a common distance of 51~kpc. It shows
similarities between their SEDs in the optical and 16--30-$\mu$m regions,
but differences in the 3--8-$\mu$m region. The day 678 clumpy model for
SN~2003gd \citep{Sug06}, required a dust mass that is 13 times higher than
the maximum mass allowed for mixed graphite:silicate models of SN~1987A
and 27 times higher than our day 775 Clumpy~{\sc ii} pure graphite mass
estimate for SN~1987A. The main reason for this difference is the very
different opacities of the dust mixtures required to fit the SEDs of the
two SNe; in the case of SN~2003gd, a dust mixture composed of 15\%
amorphous carbon and 85\% astronomical silicates provided the best fit
\citep{Sug06}, while the maximum silicate mass fraction allowed for
SN~1987A is only 15\%, due to its featureless mid-IR spectrum at late
epochs. Astronomical silicates are much more transparent to optical and
near-IR radiation than are amorphous carbon or graphite grains, and
therefore require larger dust masses to match a given SED. Furthermore, 
the much higher optical albedos of silicate grains allows much larger 
masses to be inside clumps, for similar levels of IR emission, though the 
silicate 10-$\mu$m feature should be present in emission or absorption 
(mainly the latter). We note that models containing only carbon grains did 
not provide a good fit to the day 499 optical-IR spectrum of SN~2003gd, 
producing too much emission in the 2--10-$\mu$m region. 

The model fits imply that the proportions of silicates and carbon grains
were reversed in SN~1987A and SN~2003gd. The mass of the immediate
progenitor of SN~2003gd is estimated to have been
8$^{+4}_{-2}$~M~$_{\odot}$ \citep{VDy03,Sma04}, while the mass of
Sk~--69~202, the progenitor of SN~1987A, is estimated to have been
$\sim$20~M$_{\odot}$ \citep{Arn89}. Nucleosynthetic calculations show that
although large masses of oxygen should be produced by high mass
supernovae, below 20~M~$_{\odot}$ the yield of oxygen from the O-rich zone
can drop below the yield of carbon from the C-rich zone \citep{Arn89,
  Woo95},
although the results are highly sensitive to details of some of the
adopted reaction rates \citep{Arn96}. Irrespective of this, \citet{Cla99}
and \citet{Den06} have shown that due to the destruction by ionizing
particles of CO, and the consequent disruption of the normal CO sink for
carbon atoms in oxygen-rich regions, large quantities of carbon-rich dust
particles can form easily in SN ejecta zones that have C/O$<$1. What
remains to be elucidated is what causes either silicate or carbon grains
to dominate the dust produced by particular supernovae.

\section{Discussion}

We have studied the effects of a number of clumping parameters on the SEDs 
emerging from two geometrical configurations, consisting of 
a spherical shell of clumpy medium illuminated by a diffuse radiation source.
Clumpy~{\sc i} models assume full mixing between the dust and the 
radiating medium, while Clumpy~{\sc ii} models assume that 
dust is only present in the clumps and the illuminating radiation comes 
only from the ICM.

The main conclusions that can can be drawn from our parameter investigation 
can be summarised as follows: 

\begin{enumerate}
\item varying the volume filling factor, $f$ has opposite 
effects in Clumpy~{\sc i} and Clumpy~{\sc ii} models. For a constant dust mass, 
$M_{\rm d}$ and clump to ICM ratio, $\alpha$, an increase in $f$ in Clumpy~{\sc i} 
models causes a decrease in the effective optical depth, with less optical radiation
being reprocessed at IR wavelengths. In Clumpy~{\sc ii} models $f$ directly 
controls the amount of optical radiation being intercepted by the dust, since the illumination 
source is always external of the clumps. This is the same behaviour shown by 
centrally illuminated shell models.  
\item in the range explored, variations of $\alpha$ in 
Clumpy~{\sc i} models (constant $M_{\rm d}$ and constant $f$)  have only a
 minor impact on the emerging SEDs. 
\item for models where the emitting source is intermixed with 
the dusty medium, dust grains cannot be hidden in optically thick clumps. For these models 
a degeneracy exists between $M_{\rm d}$ and $f$ which may not be resolved with IR SED 
observations alone. 
\item  for Clumpy~{\sc ii} models with optically thick clumps,
an increase in $M_{\rm d}$ results in cooler grains which will therefore radiate 
at longer IR wavelengths.
\end{enumerate}

We used both smoothly varying and clumped dust density distributions to obtain new 
estimates for the mass of dust condensed by the Type~II SN~1987A by fitting 
the 0.3-30$\mu$m spectrophotometric data of \citet{Woo93} at two epochs (day 615 and day 775). 
From our modelling we conclude that between 2$\times10^{-4}$ and 1.3$\times10^{-3}$~M$_{\odot}$ 
of mainly carbon-based grains had formed in the ejecta by day 615 (4--7$\times10^{-4}$~M$_{\odot}$)
for pure graphite models) with the derived dust masses being
found to be independent of whether the dust was clumped or not.  This
result differs from the conclusions of W93, who derived unclumped graphite
dust masses of 4--6$\times10^{-5}$~M$_{\odot}$ on days 615 and 775, versus
graphite masses from their analytic clumped dust modelling of
3--5$\times10^{-4}$~M$_{\odot}$. The latter values are close to our own
clumped graphite dust mass estimates from Monte Carlo modelling.
For a 20~M$_{\odot}$ SN progenitor, between 0.3--0.4~M$_{\odot}$ of
refractory elements are expected to form (Woosley \& Weaver 1995), so the
estimate of 2$\times10^{-4}$ to 1.3$\times10^{-3}$~M$_{\odot}$ of new dust
formed in the ejecta of SN~1987A corresponds to a condensation efficiency
of only 5$\times10^{-4}$ to 4$\times10^{-3}$, significantly lower than the
dust condensation efficiency of $\sim$0.12 estimated for SN~2003gd by
Sugerman et al. (2006). The cause of such very different dust yields
amongst Type~II core-collapse supernovae remains to be explained.

{\bf Acknowledgments: }
We are grateful to Dr. P. Bouchet, the referee, for helpful and
constructive comment. BS acknowledges support from HST grant 10607 and SST
grant 20320.



\begin{thebibliography}{}



\bibitem[\protect\citeauthoryear{Abbott \& Lucy}{1985}]{AL85} Abbott
D.~C., Lucy L.~B., 1985, ApJ, 288, 679

\bibitem[\protect\citeauthoryear{Arnett}{1996}]{Arn96} Arnett, D.\
  1996, Supernovae and Nucleosynthesis, Princeton: Princeton
  University Press, p.\ 302

\bibitem[\protect\citeauthoryear{Arnett et al.}{1989}]{Arn89} Arnett
  W.~D., Bahcall J.~N., Kirshner R.~P., Woosley S.~E., 1989, ARA\&A,
  27, 629

\bibitem[\protect\citeauthoryear{Bertoldi et al.}{2003}]{Ber03}
Bertoldi F., Carilli C.~L., Cox P., Fan X., Strauss M.~A., Beelen A.,
Omont A., Zylka R., 2003, A\&A, 406, L55

\bibitem[\protect\citeauthoryear{Boisse}{1990}]{Boi90} Boisse P.,
  1990, A\&A, 228, 483

\bibitem[\protect\citeauthoryear{Bouchet}{2004}]{Bou04} Bouchet, P.,
  De Buizer, J. M., Suntzeff, N. B., Danziger, I. J., Hayward, T. L., 
  Telesco, C. M., Packham, C., 2004, ApJ, 611, 394

\bibitem[\protect\citeauthoryear{Bouchet \& Danziger}{1993}]{BD93}
Bouchet P., Danziger I.~J., Lucy L.~B., 1991, AJ, 102, 1135

\bibitem[\protect\citeauthoryear{Bouchet \& Danziger}{1991}]{Bou91}
Bouchet P., Danziger I.~J., 1993, A\&A, 273, 451

\bibitem[\protect\citeauthoryear{Caldwell et al.}{1993}]{Cal93} 
Caldwell et al., 1993, MNRAS, 262, 313

\bibitem[\protect\citeauthoryear{Cernuschi, Marsicano, \&
Kimel}{1965}]{Cer65} Cernuschi F., Marsicano F.~R., Kimel I., 1965,
AnAp, 28, 860

\bibitem[\protect\citeauthoryear{Chevalier \& Klein}{1978}]{CK78}
Chevalier R.~A., Klein R.~I., 1978, ApJ, 219, 994

\bibitem[\protect\citeauthoryear{Chevalier \& Emmering}{1989}]{CE89}
Chevalier R.~A., Emmering R.~T., 1989, ApJ, 342, L75

\bibitem[\protect\citeauthoryear{Clayton, Amari \& Zinner}{1997}]{Cla97}
Clayton D.~D., Amari S., Zinner E., 1997, ApSS, 251, 355

\bibitem[\protect\citeauthoryear{Clayton, Liu \& Dalgarno}{1999}]{Cla99}
Clayton D.~D., Liu W., Dalgarno A., 1999, Science, 283, 1290

\bibitem[\protect\citeauthoryear{Deneault, Clayton \& Heger}{2006}]{Den06}
Deneault E.~A.-N., Clayton D.~D., Heger A., 2006, ApJ, 594, 312

\bibitem[\protect\citeauthoryear{Douvion, Lagage \& Cesarsky}{1999}]{Dou99} 
Douvion T., Lagage P.~O., Cesarsky C.~J., 1999, A\&A, 352, L111

\bibitem[\protect\citeauthoryear{Draine \& Lee}{1984}]{DL84} 
Draine B.~T., Lee H.~M., 1984, ApJ, 285, 89 

\bibitem[\protect\citeauthoryear{Dwek}{1988}]{Dwe88} Dwek E., 1988,
  ApJ, 329, 814

\bibitem[\protect\citeauthoryear{Dwek et al.}{1992}]{Dwe92} Dwek E.,
  Moseley S.~H., Glaccum W., Graham J.~R., Loewenstein R.~F.,
  Silverberg R.~F., Smith R.~K., 1992, ApJ, 389, L21

\bibitem[\protect\citeauthoryear{Elmhamdi et al.}{2003}]{Elm03}
Elmhamdi A., et al., 2003, MNRAS, 338, 939


\bibitem[\protect\citeauthoryear{Elvis, Marengo \&
Karovska}{2002}]{Elv02} Elvis M., Marengo M., Karovska M., 2002, ApJ,
567, L107

\bibitem[\protect\citeauthoryear{Ercolano et al.}{2003a}]{Erc03}
Ercolano B., Barlow M.~J., Storey P.~J., Liu X.-W., 2003, MNRAS, 340,
1136

\bibitem[\protect\citeauthoryear{Ercolano et al.}{2003b}]{Erc03b}
Ercolano B., Morisset C., Barlow M.~J., Storey P.~J., Liu X.-W., 2003, 
MNRAS, 340, 1153

\bibitem[\protect\citeauthoryear{Ercolano, Barlow \&
Storey}{2005}]{Erc05} Ercolano B., Barlow M.~J., Storey P.~J., 2005,
MNRAS, 362, 1038

\bibitem[\protect\citeauthoryear{Gordon, Calzetti \&
Witt}{1997}]{Gor97} Gordon K.~D., Calzetti D., Witt A.~N., 1997, ApJ,
487, 625

\bibitem[\protect\citeauthoryear{Hamuy \& Suntzeff}{1990}]{Ham90}
Hamuy M., Suntzeff N. B., 1990, AJ, 99, 1146

\bibitem[\protect\citeauthoryear{Hanner}{1988}]{Han88}Hanner M. S.,
1988, NASA Conf. Publ., 3004, 22

\bibitem[\protect\citeauthoryear{Harvey et al.}{1989}]{Har89} Harvey
    P., Lester D., Dinerstein H., Smith B., Colome C., 1989, BAAS, 21,
    1215

\bibitem[\protect\citeauthoryear{Herant \& Woosley}{1994}]{HW94}
Herant M., Woosley S.~E., 1994, ApJ, 425, 814

\bibitem[\protect\citeauthoryear{Hines et al.}{2004}]{Hin04} Hines
    D.~C., et al., 2004, ApJS, 154, 290

\bibitem[\protect\citeauthoryear{Hirashita et al.}{2005}]{Hir05}
Hirashita H., Nozawa T., Kozasa T., Ishii T.~T., Takeuchi T.~T., 2005,
MNRAS, 357, 1077

\bibitem[\protect\citeauthoryear{Hobson \& Padman}{1993}]{HP93}
Hobson M.~P., Padman R., 1993, MNRAS, 264, 161

\bibitem[\protect\citeauthoryear{Hobson \& Scheuer}{1993}]{HS93}
Hobson M.~P., Scheuer P.~A.~G., 1993, MNRAS, 264, 145


\bibitem[\protect\citeauthoryear{Kozasa, Hasegawa \&
Nomoto}{1991}]{Koz91} Kozasa T., Hasegawa H., Nomoto K., 1991, A\&A,
249, 474

\bibitem[\protect\citeauthoryear{Krause et al.}{2004}]{Kra04} Krause
    O., Birkmann S.~M., Rieke G.~H., Lemke D., Klaas U., Hines D.~C.,
    Gordon K.~D., 2004, Natur, 432, 596

\bibitem[\protect\citeauthoryear{Lucy et al.}{1989}]{Luc89}Lucy L. B.,
Danziger I. J., Gouiffes C., Bouchet P., 1989, in IAU Colloq. 120,
Structure and dynamics of the interstellar medium,
ed. G. Tenorio-Tagle, M. Moles, J. Melnick, Springer-Verlag, 164

\bibitem[\protect\citeauthoryear{Lucy et al.}{1991}]{Luc91} Lucy
    L. B., Danziger I. J., Gouiffes C., Bouchet P., 1991, in
    Supernovae, ed. S. E. Woosley, Springer-Verlag, 82


\bibitem[\protect\citeauthoryear{Maiolino et al.}{2004}]{Mai04}
Maiolino R., Schneider R., Oliva E., Bianchi S., Ferrara A., Mannucci
F., Pedani M., Roca Sogorb M., 2004, Natur, 431, 533

\bibitem[\protect\citeauthoryear{Mathis}{1990}]{Mat90} Mathis J.~S.,
  1990, ARA\&A, 28, 37

\bibitem[\protect\citeauthoryear{Mathis, Rumpl \&
Nordsieck}{1977}]{MRN77} Mathis J.~S., Rumpl W., Nordsieck K.~H.,
1977, ApJ, 217, 425

\bibitem[\protect\citeauthoryear{Morgan \& Edmunds}{2003}]{ME03}
Morgan H.~L., Edmunds M.~G., 2003, MNRAS, 343, 427
    
\bibitem[\protect\citeauthoryear{Moseley et al.}{1989}]{Mos89} 
, 1989, ApJ, 347, 1119

\bibitem[\protect\citeauthoryear{Natta \& Panagia}{1984}]{Nat84} Natta
A., Panagia N., 1984, ApJ, 287, 228

\bibitem[\protect\citeauthoryear{Neufeld}{1991}]{Neu91} Neufeld D.~A.,
1991, ApJ, 370, L85

\bibitem[\protect\citeauthoryear{Robson et al.}{2004}]{Rob04} Robson
  I., Priddey R.~S., Isaak K.~G., McMahon R.~G., 2004, MNRAS, 351, L29

\bibitem[\protect\citeauthoryear{Roche, Aitken \&
Smith}{1993}]{Roc93} Roche P.~F., Aitken D.~K., Smith C.~H., 1993,
MNRAS, 261, 522

\bibitem[\protect\citeauthoryear{Scuderi et al.}{1996}]{Scu96} Scuderi
S., Panagia N., Gilmozzi R., Challis P.~M., Kirshner R.~P., 1996, ApJ,
465, 956

\bibitem[\protect\citeauthoryear{Smartt et al.}{2004}]{Sma04} Smartt
    S.~J., Maund J.~R., Hendry M.~A., Tout C.~A., Gilmore G.~F.,
    Mattila S., Benn C.~R., 2004, Sci, 303, 499

\bibitem[\protect\citeauthoryear{Sugerman et al.}{2006}]{Sug06}
Sugerman B.~E.~K., et al., 2006, Sci, 313, 196

\bibitem[\protect\citeauthoryear{Suntzeff \& Bouchet}{1990}]{SB90}
  Suntzeff N.~B., Bouchet P., 1990, AJ, 99, 650

\bibitem[\protect\citeauthoryear{Todini \& Ferrara}{2001}]{TF01}
Todini P., Ferrara A., 2001, MNRAS, 325, 726

\bibitem[\protect\citeauthoryear{Travaglio et al.}{1999}]{Tra99}
Travaglio C., Gallino R., Amari S., Zinner E., Woosley S., Lewis R.~S.,
1999, ApJ, 510, 325

\bibitem[\protect\citeauthoryear{Van Dyk, Li, \&
Filippenko}{2003}]{VDy03} Van Dyk S.~D., Li W., Filippenko A.~V.,
2003, PASP, 115, 1289

\bibitem[\protect\citeauthoryear{V{\'a}rosi \& Dwek}{1999}]{VD99}
V{\'a}rosi F., Dwek E., 1999, ApJ, 523, 265

\bibitem[\protect\citeauthoryear{Weingartner \& Draine}{2001}]{WD01} 
Weingartner J.~C., Draine B~T., 2001, ApJ, 548, 296

\bibitem[\protect\citeauthoryear{Witt \& Gordon}{1996}]{Wit96} Witt
A.~N., Gordon K.~D., 1996, ApJ, 463, 681

\bibitem[\protect\citeauthoryear{Wolf, Fischer, \& Pfau}{1998}]{Wol98}
Wolf S., Fischer O., Pfau W., 1998, A\&A, 340, 103

\bibitem[\protect\citeauthoryear{Wooden et al.}{1993}]{Woo93} Wooden
    D.~H., Rank D.~M., Bregman J.~D., Witteborn F.~C., Tielens
    A.~G.~G.~M., Cohen M., Pinto P.~A., Axelrod T.~S., 1993, ApJS, 88,
    477

\bibitem[\protect\citeauthoryear{Woosley \& Weaver}{1995}]{Woo95}
Woosley S.~E., Weaver T.~A., 1995, ApJS, 101, 181

\bibitem[\protect\citeauthoryear{Woosley et al.}{1989}]{Woo89}
  Woosley S.~E., Hartmann D., Pinto P.~A., 1989, ApJ, 346, 395

\end{thebibliography}
\end{document}